\documentclass[reprint,prd,12pt,superscriptaddress,onecolumn]{revtex4-2}
\usepackage{hyperref}
\usepackage{ifpdf}
\usepackage{subfig}
\usepackage[normalem]{ulem}
\usepackage{amsmath,amssymb,amsfonts,gensymb}
\usepackage{epsf}
\usepackage{rotating}
\usepackage{nicefrac}
\usepackage{color,xcolor,pstricks}
\usepackage{fancyhdr}
\usepackage{lineno}
\usepackage{braket}
\usepackage{multirow}
\usepackage{empheq}
\usepackage{graphics}

\definecolor{bv}{rgb}{0.54, 0.17, 0.89}
\allowdisplaybreaks

\def\bar {\overline}

\def\delacp{\Delta A_{\rm CP}}

\def\beq{\begin{equation}}
\def\eeq{\end{equation}}
\def\bea{\begin{eqnarray}}
\def\eea{\end{eqnarray}}
\def\barr{\begin{array}}
\def\earr{\end{array}}
\def\nn {\nonumber}

 \usepackage{color}

 \usepackage[normalem]{ulem}

 \definecolor{darkgreen}{cmyk}{1,0,1,0.4}

\def\com2#1{\textcolor{red}{\textit{#1}}}

\begin{document}

\title{Complete analysis of all $B\to \pi K$ decays}

\author{Anirban Kundu}
\email{akphy@caluniv.ac.in}
\affiliation{Department of Physics, University of Calcutta, 92 Acharya Prafulla Chandra Road, Kolkata 700009, India}

\author{Sunando Kumar Patra}
\email{sunando.patra@gmail.com}
\affiliation{Department of Physics, Bangabasi Evening College, 19 Rajkumar Chakraborty Sarani, Kolkata, 700009, West Bengal, India}
	
\author{Shibasis Roy} 
\email{shibasis.cmi@gmail.com}
\affiliation{Department of Physics, University of Calcutta, 92 Acharya Prafulla Chandra Road, Kolkata 700009, India}

\begin{abstract}
The Standard Model (SM) predicts that $\Delta A_{\rm CP}$, the difference between the direct CP asymmetries 
for the modes $B^+\to \pi^0 K^+$ and $B^0\to \pi^- K^+$ that are related by weak isospin, should be close to zero. 
There has been a recent claim by the LHCb Collaboration that the measured value of $\Delta A_{\rm CP}$ shows an 
uncomfortable tension with the SM prediction, almost at the $8\sigma$ level. Motivated by this claim, we critically 
re-examine the data on all the $B\to \pi K$ modes, including the CP asymmetries and CP-averaged branching fractions. 
From a combined Bayesian analysis with the topological amplitudes and their phases as the free parameters, we find that 
the best-fit region has a large overlap with the parameter space favoured in the SM, albeit with some enhancement for the electroweak penguin and the colour-suppressed tree amplitudes, consistent with the findings of 
earlier studies. We find that in this SM-like region, $\Delta A_{\rm CP}$ is more than 5$\sigma$ away from zero 
and the tension with the global average, as well as the LHCb result, is within $2\sigma$. Thus we conclude that
there is not yet enough motivation to go beyond the SM. 
\end{abstract}
 
\maketitle

\section{Introduction}
   \label{sec:intro}
 
The neutral and charged $B$-mesons decaying to a $\pi K$ pair, namely, $B^+\to \pi^0 K^+$, $\pi^+ K^0$, and 
$B^0\to \pi^- K^+$, $\pi^0 K^0$ (plus the CP-conjugate channels), continue to show some tension when the four branching 
ratios (BR), four direct CP asymmetries and the mixing-induced CP-asymmetry in $B^{0}\to \pi^{0}K^{0}$ are compared with 
the Standard Model (SM) predictions. This is, in essence, the ``$B\to \pi K$ 
puzzle"~\cite{Buras:2003yc,Buras:2003dj,Buras:2004ub,Baek:2004rp}; the experimental data from the 
BaBar~\cite{Aubert:2006fha,Aubert:2006gm,Aubert:2007hh,Aubert:2008ad,Lees:2012mma},
Belle~\cite{Fujikawa:2008pk,Duh:2012ie}, LHCb~\cite{Aaij:2013fja,Aaij:2018tfw,Aaij:2020buf}, 
and very recently, Belle-II~\cite{Abudinen:2021gly,Abudinen:2021zmr,Abudinen:2021dyj} Collaborations have continued to provide 
support for the puzzle.
Recently, the LHCb Collaboration have updated the data on the difference of the direct CP asymmetries~\cite{Aaij:2020wnj}
between the modes $B^+\to \pi^0 K^+$ and $B^0\to \pi^- K^+$, defined as $\delacp$:
\beq
\delacp^\text{LHCb}(\pi K)=0.108\pm 0.017
\eeq
that is non-zero with a significance of more than 6 standard deviations while the global average
\beq
\delacp^\text{global}(\pi K)=0.115\pm 0.014
\eeq
lies a remarkable 8 standard deviations away from zero, which happens to be the SM prediction~\cite{Gronau:1998ep} 
for $\delacp$. The null prediction of the SM is based on the relative importance of certain flavour-flow 
topologies~\cite{Gronau:1998ep} as well as on the assumption that the difference in strong phase between the tree and 
electroweak penguin amplitudes is identically zero. The goal of this paper
is to check the robustness of these claims.

A good way of writing these topological amplitudes is shown in Ref.\ \cite{Gronau:1994rj,Gronau:1995hn,Gronau:1995hm}, which we follow. We first enlist all 
the relevant amplitudes~\cite{Gronau:1995hn}, including the possible weak and strong phases. The details can be found in Section 
\ref{sec:groundwork}. We implicitly assume that the weak phase comes 
solely from the Cabibbo-Kobayashi-Maskawa (CKM) mixing matrix elements, while the strong phases can be arbitrary. 
The latter come mostly from long-distance re-scattering effects ~\cite{Neubert:1997wb,Atwood:1997iw,Buras:1997cv,Buras:2000gc,Falk:1998wc}, so one may theoretically expect some relationship 
among the strong phases~\cite{Buras:1998ra,Beneke:2000ry,Bauer:2004ck,Bauer:2004tj,Bauer:2005kd,Beneke:2001ev,Huitu:2009st} associated with various amplitudes (up to a certain level of precision), but we try to avoid 
such theoretical prejudices as much as possible. 
The analysis treats all the amplitudes and relevant strong phases as free parameters, while the CKM elements and weak
phases are incorporated as multi-normal priors. The numbers on any specific observable coming from different experiments 
are treated as independent inputs in the absence of any correlation, {\em i.e.}, we do not take the average values for the 
observables. This, naturally, enhances the number of data points. For the justification of this approach, and 
for the methodology of the analysis, we refer the reader to Section \ref{sec:methodology}.

What we find is not completely unexpected in view of the previous analyses~\cite{Nandi:2004dx,Chang:2005jn,Baek:2006ti,Baek:2007yy,Kim:2007kx,Feldmann:2008fb,Ciuchini:2008eh,Baek:2009pa,Baek:2009hv,Beaudry:2017gtw,Fleischer:2017vrb,Fleischer:2018bld}, but
sheds some interesting light on the so-called puzzle. The salient features, discussed in more detail in Section \ref{sec:results}, are 
as follows. 

(i) We performed a Bayesian analysis with uniform priors on the topological amplitudes and phases supplied over a 
wide range. 
We have also checked that this region contains the $\chi^2$-minimum of the frequentist analysis.

(ii) With a naive estimate of the relative importance of the amplitudes~\cite{Gronau:1994rj,Gronau:1995hn,Imbeault:2005ne}, 
there is no acceptable fit to the data. This estimate 
is mostly based on the CKM factors present in each amplitude; the smallness is controlled by $\lambda$, the sine of the 
Cabibbo angle ($\sim 0.22$). 
Another important parameter that comes into play is the ratio of the Wilson Coefficients (WC) of the electroweak 
penguin and tree operators. 
There is no reason why the low-energy QCD corrections, 
including the effects coming from the running of the WCs of the relevant operators, should still 
respect these estimates. Thus, based on this part of the analysis, one must not say that the SM is ruled out. 

(iii) One might like to relax the bounds on the colour-suppressed tree amplitudes to cover possible non-perturbative QCD effects~\cite{Buras:1998ra,Beneke:2000ry,Keum:2000wi,Beneke:2003zv,Beneke:2001ev,Chang:2008tf,Cheng:2009eg,Li:2009wba,Li:2014haa,Liu:2015upa} as well as the electroweak penguin amplitudes.

This, in effect, should cover the entire SM-allowed region, but some of the amplitudes may be enhanced
compared to the naive fit. As such enhancements are not ruled out even within the framework of the SM, 
this is what we call the `SM-like' region, more details of which are given later. 
However, one may also invoke new physics (NP) to explain this~\cite{Datta:2004re,Imbeault:2006nx,Hofer:2010ee,Crivellin:2019isj,Calibbi:2019lvs,Bhattacharya:2021ggm}. 
In this SM-like parameter space with possibilities of NP playing a role
\cite{Baek:2006ti,Baek:2007yy,Kim:2007kx,Feldmann:2008fb,Ciuchini:2008eh,Baek:2009pa,Baek:2009hv,Beaudry:2017gtw,Datta:2019tuj} lies the best-fit region, with $\delacp$ more than $5\sigma$ away from zero, and with a tension of
less than $2\sigma$ with the global average. This leads to our main conclusion:
the data is still not at variance with the SM, and the discrepancy in $\delacp$ is not something to claim the existence of
NP. However, presence of NP cannot be ruled out either. 

(iv) Finally, we entertain the possibility of NP in $B\to \pi K$, which might affect the amplitudes, as well as the 
WCs. We perform a free fit, relaxing the SM limit on the amplitudes. 
In the SM, the ratio of the electroweak penguin amplitude to the tree amplitude is approximately the same,
for both colour-allowed and colour-suppressed channels. In the free fit, these two ratios may also possibly differ. 
We find that there appears another best-fit region, but definitely beyond the SM-allowed parameter space.

One might ask whether there is any way to differentiate between these two best-fit regions and know for sure which
is the actual one. We show that there are two possible ways. The SM-like prediction for $\delacp$ is about $2\sigma$ 
away from the global average; if the data becomes more precise and the tension increases, it might point to the presence 
of NP. Alternatively, a suitably defined combination of the branching ratios and CP asymmetries of
these four channels may also have the potential to do the job.

The paper, thus, is arranged as follows. In Section \ref{sec:groundwork}, we display the relevant expressions, and lay out 
the relevant parameter spaces. Section \ref{sec:methodology} is about the analysis, 
while we show our results in Section \ref{sec:results}. Section \ref{sec:summary} summarises and concludes the paper.  

\section{Theory inputs}
\subsection{Topological amplitudes and phases}   
\label{sec:groundwork}
\begin{figure}
	\includegraphics[width=0.8\textwidth]{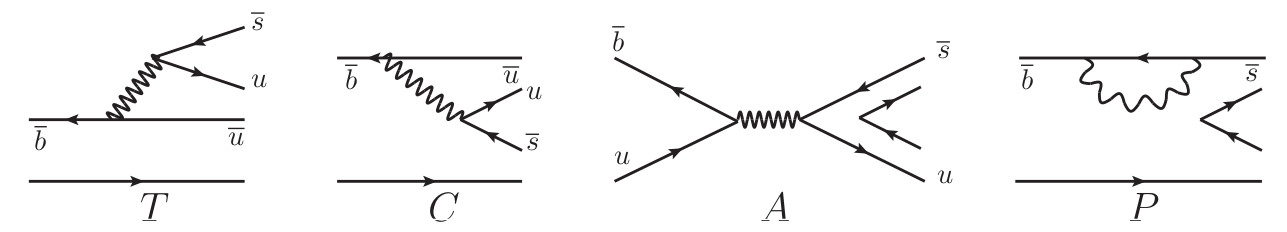}
	\caption{\small Topological amplitudes for $B\to \pi K$ decays, taken from Ref.~\cite{Gronau:1994rj}.}
	\label{fig:diagram1}
\end{figure}
The four $B\to  \pi K$ decay amplitudes can be parametrised in at least two different ways. The first one is 
by three isospin amplitudes~\cite{Nir:1991cu,Lipkin:1991st,Gronau:1998ep,Neubert:1998re} $A_{1/2}$, $A_{3/2}$, and $B_{1/2}$, 
where $A$ and $B$ respectively stand for $\Delta I=1$ and $\Delta I=0$ amplitudes, and the subscript denotes the 
isospin of the $\pi K$ combination. In this notation, the amplitudes are written as~\cite{Neubert:1998re} 
\begin{eqnarray}
	A(B^+\to \pi^+K^0) &=& B_{1/2} + A_{1/2} + A_{3/2}\,,\nn\\
	A(B^+\to \pi^0K^+) &=& -\frac{1}{\sqrt{2}}\left( B_{1/2} + A_{1/2}\right) + \sqrt{2} A_{3/2}\,,\nn\\
	A(B^0\to \pi^-K^+) &=& - B_{1/2} + A_{1/2} + A_{3/2}\,,\nn\\
	A(B^0\to \pi^0K^0) &=& \frac{1}{\sqrt{2}}\left( B_{1/2} - A_{1/2}\right) + \sqrt{2} A_{3/2}\,.
\end{eqnarray}
The QCD penguin diagram contributes only to the isospin amplitude $B_{1/2}$ and hence it is the largest among 
the three. Each such amplitude is actually made of two parts, with two independent CKM factors. Thus, one has six independent 
amplitudes and therefore five independent strong phase differences. 

The hierarchy among the amplitudes is even more evident if we consider the topological amplitudes~\cite{Fleischer:1997um,Gronau:1997an,Neubert:1998pt,Neubert:1998jq,Gronau:2006xu}. There are four
distinct topological classes, namely, ``colour-allowed tree" $T$, ``colour-suppressed tree" $C$, ``annihilation" $A$, and 
``penguin". These topologies are shown in Fig.\ \ref{fig:diagram1}, taken from~\cite{Gronau:1994rj,Gronau:1995hn}.
The penguin topology is further subdivided into strong penguin $P$, colour-allowed electroweak penguin
(EWP) $P_{\rm EW}$, and colour-suppressed EWP $P^C_{\rm EW}$. Each amplitude carries its own strong phase, but only
the phase differences are relevant, as an overall phase does not have any effect on the observables. 

The decay amplitudes for $B^0\to \pi^-K^+$, $B^+\to \pi^+ K^0$, $B^0\to \pi^0 K^0$, and $B^+\to \pi^0 K^+$,
from now on denoted by $\mathcal{A}^{-+}$, $\mathcal{A}^{+0}$, $\mathcal{A}^{00}$, $\mathcal{A}^{0+}$ respectively, may be 
expressed in terms of the topological amplitudes~\cite{Chau:1990ay,Gronau:1994rj,Gronau:1995hn,Gronau:2005kz} as
\begin{eqnarray}
	\mathcal{A}^{-+} &=& -\lambda_{u} \left(P_{uc}+T\right) - \lambda_{t} \left(P_{tc}+\frac23 P^{C}_{EW}\right)\,,\nn\\
	\mathcal{A}^{+0} &=& \lambda_{u} \left(P_{uc}+A \right)+\lambda_{t} \left(P_{tc}-\frac{1}{3}P^{C}_{EW}\right)\,,\nn\\
	\sqrt{2}\mathcal{A}^{00} &=& \lambda_{u} \left(P_{uc}-C\right) + \lambda_{t} \left(P_{tc}-P_{EW}-\frac{1}{3}P^{C}_{EW}\right)
	\,,\nn\\ 
	\sqrt{2}\mathcal{A}^{0+} &=&-\lambda_{u} \left(P_{uc}+T+C+A\right) - \lambda_{t} \left(P_{tc}+P_{EW}+\frac{2}{3}P^{C}_{EW}
	\right)\,,\label{Bto0P}
\end{eqnarray}

where we have factored out the CKM elements $\lambda_{q}=V_{qb}^{*}V_{qs}$ from the amplitudes. 
We note that the penguin amplitude $P$ receives contributions from all the three up-type quarks in the loop:
\begin{align}
	P = \lambda_{u}P_{u} + \lambda_{c}P_{c} + \lambda_{t}P_{t}\,,
\end{align}
and the unitarity of the CKM matrix
\begin{align}
	\lambda_{u} + \lambda_{c} + \lambda_{t} = 0
\end{align}
leads to
\begin{align}
	P=\lambda_{u} \left( P_{u} - P_{c} \right) + \lambda_{t} \left( P_{t} - P_{c} \right) \equiv \lambda_u P_{uc} + \lambda_t P_{tc}\,.
\end{align}

We expect a hierarchy among these amplitudes, which, magnitude-wise, looks like
\begin{align}
\left|\lambda_{t}P_{tc}\right| \, > \, \left|\lambda_{u}T\right|\, > \, 
\left|\lambda_{u}C\right| \, > \, \left|\lambda_{u}A\right|\,,\,  \left|\lambda_{u}P_{uc}\right|\,,
\label{eq:hier}
\end{align}
where every subsequent step is suppressed compared to the previous one by a factor of the order of 
$\lambda \approx \sin\theta_C = 0.22$, $\theta_C$ being the Cabibbo angle. 
The suppression is a combined effect of the magnitudes of the respective CKM elements and the extra loop 
suppression of the penguin amplitudes. For example, $\lambda_u/\lambda_t \sim \lambda^2$, but $P_{tc}$ 
is loop-suppressed compared to $T$, again by an order of $\lambda$~\cite{Gronau:1995hm}.
It also turns out that $|C/T| \sim \lambda$~\cite{Beneke:2001ev}. 
However, even within the SM, $|C/T| \sim 0.5$ is definitely possible~\cite{Li:2009wba}, which we will 
use in our analysis. One may note that this ratio of the order of unity~\cite{Bauer:2005kd,Huitu:2009st} is also not ruled out. 
The annihilation amplitude $A$ is suppressed by a factor of $f_{B}/m_{B}\sim 0.05 \sim \lambda^2$ 
when compared to $T$. The long-distance re-scattering 
effects should modify these predictions, but if the colour-transparency argument holds for $B\to\pi K$, we do not
expect a drastic reordering. Thus, anything widely off from Eq.\ (\ref{eq:hier}) signals the presence of NP. On the other hand,
one must remain open to the possibility that the hierarchy may not be all that sacrosanct; an enhancement by 
${\cal O}(1/\lambda) \sim 5$ for some cases even within the framework of SM may not be discarded offhand. 

A relation between the tree and the EWP amplitudes may be obtained with the help of $SU(3)$-flavour symmetry 
of the dimension-6 weak Hamiltonian mediating the $\vert \Delta S\vert$=1 decay~\cite{Buchalla:1995vs,Gronau:1998fn},
\begin{align}
	\mathcal{H}\left( \Delta S=1 \right) = \frac{G_{F}}{\sqrt{2}} \left[\lambda_{u}\left( C_{1} \left(\overline{b}u \right)_{V-A}
	\left(\overline{u}s \right)_{V-A} + C_{2} \left( \overline{b}s \right)_{V-A} \left( \overline{u}u \right)_{V-A} \right) -\lambda_{t} \sum_{i=3}^{10} C_{i} Q_{i} \right]
\end{align}
where $Q_{1-2}$ are the tree, $Q_{3-6}$ are the QCD penguin and $Q_{9},\,Q_{10}$ are the two 
non-negligible EWP operators in the SM. The $P_{EW}$ and $P^{C}_{EW}$ amplitudes are therefore given in terms of 
$T$ and $C$ respectively~\cite{Neubert:1998jq,Neubert:1998pt,Gronau:1998fn},
\beq
P_{EW} \pm P_{EW}^{C}=-\frac{3}{2}\, \frac{C_{9} \pm C_{10}} {C_{1} \pm C_{2}}\, (T \pm C)\,.
\eeq
After plugging in the numerical values of the WCs $C_1$, $C_2$,
$C_9$, and $C_{10}$ to the leading log order at $m_{b}$ scale~\cite{Buchalla:1995vs}, one gets
\beq
\label{T-PEW reln}
P_{EW}\sim \kappa T\,,\ \ \qquad
P_{EW}^{C}\sim \kappa C\,,
\eeq
to a good approximation, where 
\beq
\kappa=-\frac{3}{2}\, \frac{C_{9}+C_{10}}{C_{1}+C_{2}}\simeq -\frac{3}{2}\, \frac{C_{9}-C_{10}}{C_{1}-C_{2}}\simeq 
0.0135\pm 0.0012\,.
\label{eq:kappa-def}
\eeq 
One may note that in SM, both $P_{EW}/T$ and $P^C_{EW}/C$ are approximately the same. This need not be true in the
presence of NP, or even some yet-to-be-accounted for SM dynamics.
As the $\kappa$-suppression compensates the $\lambda_t/\lambda_u$ enhancement, one may infer that 
$\vert \lambda_{t} P_{EW}\vert \sim \vert\lambda_u T\vert$ and 
$\vert \lambda_{t} P_{EW}^{C}\vert\sim \vert \lambda_u C\vert$. Thus, these two amplitudes may have a non-negligible contribution to branching fractions and CP asymmetries for $B\to \pi K$ decays.

It is worth mentioning here that any NP contributing to $P_{EW}$ simultaneously affects $C$ as well, as their 
respective contributions enter the decay amplitudes in Eq.\ \eqref{Bto0P} exclusively in a particular combination~\cite{Imbeault:2006nx}. 
This attribute is not limited to the above mentioned combination of diagrammatic amplitudes, but is an artifact of redundancy in the 
definition of them. This is known as reparametrisation invariance, and prevents a clean extraction of NP affecting a 
particular amplitude from the available experimental observations.

\subsection{CP asymmetries}\label{sec:asymmdef}

The decay rate asymmetry for any $B\to \pi K$ process is defined as 
\begin{align}
	\Delta(\pi K)=\Gamma(b)-\Gamma(\overline{b}) 
\end{align}
where $\Gamma(b)$ and $\Gamma(\overline{b})$ are the decay rates of the CP-conjugate mesons containing a $b$-quark
({\em i.e.}, $B^-$ or $\overline{B^0}$)  or a $\bar{b}$ quark ({\em i.e.}, $B^+$ or $B^0$) respectively. 
In terms of the topological amplitudes, the four $B\to \pi K$ decay rate asymmetries~\cite{Gronau:2005kz} are given by
\begin{eqnarray}
	\Delta (\pi^- K^+) &=& -4\, \text{Im}(\lambda_{u}^{*}\lambda_{t})\, 
	\text{Im}[(T+P_{uc})^{*}(P_{tc}+\textstyle{\frac23}P^{C}_{EW})]\,,\nonumber\\
	2\Delta (\pi^0 K^+) &=& -4\, \text{Im}(\lambda_{u}^{*}\lambda_{t})\, \text{Im}[(T+C+A+P_{uc})^{*}(P_{tc}+P_{EW}
	+\textstyle{\frac{2}{3}}P^{C}_{EW})]\,,\nonumber\\
	2\Delta (\pi^0 K^0) &=& -4\, \text{Im}(\lambda_{u}^{*}\lambda_{t})\, \text{Im}[(P_{uc}-C)^{*}(P_{tc}-P_{EW}
	-\textstyle{\frac13}P^{C}_{EW})]\,,\nonumber\\
	\Delta (\pi^+ K^0) &=& -4\, \text{Im}(\lambda_{u}^{*}\lambda_{t})\, \text{Im}[(A+P_{uc})^{*}(P_{tc}
	-\textstyle{\frac13}P^{C}_{EW})]\,.
\end{eqnarray}
The direct CP asymmetry $A_{\rm{CP}}(\pi K)$ is subsequently defined as
\begin{eqnarray}
	A_{\text{CP}}(\pi K) = \frac{\Delta(\pi K)}{\Gamma(b)+\Gamma(\overline{b})}\,,
\end{eqnarray}

For $B^0$ and $\overline{B^0}$ decaying to the CP-eigenstate $f_{\rm CP}$, one may also measure the mixing-induced 
CP violation, parametrised by $S_{\rm CP}$ and defined as
\begin{align}
	A_{\text{CP}}(t)=\frac{\Gamma(\overline{B^{0}}(t)\to f)-\Gamma(B^{0}(t)\to \overline{f})}
	{\Gamma(\overline{B^{0}}(t)\to f)+\Gamma(B^{0}(t)\to \overline{f})}=
	A_{\text{CP}}(f)\cos(\delta m\, t )+S_{\text{CP}}(f)\sin(\delta m\, t )
\end{align} 
where $\delta m=m_{H}-m_{L}$ is the mass difference between the heavier and lighter $B$-meson mass 
eigenstates. 
By ignoring diagrams of $\mathcal{O}(\lambda^{2})$ and beyond, the $B\to \pi K$ amplitudes reduce to 
\begin{eqnarray}
	\mathcal{A}^{-+} &=& -\lambda_{u} T - \lambda_{t} P_{tc}\,,\nn\\
	\mathcal{A}^{+0} &=& \lambda_{t}P_{tc}\,,\nn\\
	\sqrt{2}\mathcal{A}^{00} &=&  \lambda_{t} \left(P_{tc}-P_{EW}\right)
	\,,\nn\\ 
	\sqrt{2}\mathcal{A}^{0+} &=&-\lambda_{u} T - \lambda_{t} \left(P_{tc}+P_{EW}
	\right)\,.
\end{eqnarray}
Direct CP asymmetries in $B^0\to \pi^-K^+$ and $B^+\to \pi^0 K^+$ arise because of the $T$--$P_{tc}$ interference 
leading to a non-zero relative strong phase, as well as a weak phase difference between the two topological amplitudes. 
In contrast, $P_{EW}$ and $T$ carry the same strong phase as they are related by a real number as shown in  
Eq.~\eqref{eq:kappa-def}. Therefore, $P_{EW}$--$T$ interference for $B^+\to \pi^0K^+$ does not contribute to 
$A_{\rm CP}$. Thus, one expects a simplified relation~\cite{Gronau:1998ep},
\beq
	A_{\rm CP}(B^{0}\to \pi^{-}K^{+})=A_{\rm CP}(B^{+}\to \pi^{0}K^{+}).
\eeq
Any deviation, numerically expressed by the quantity $\delacp$;
\begin{align}\label{eq:delACP}
\Delta A_{\rm CP}=A_{\rm CP}(B^{+}\to \pi^{0}K^{+})-A_{\rm CP}(B^{0}\to \pi^{-}K^{+}),
\end{align}
thus, would necessarily mean that amplitudes like $C$ may not be neglected, but it would be very premature
to claim this as a telltale signature of NP. 
A more robust CP sum rule relation~\cite{Gronau:2005kz} connecting all the four $B \to \pi K$ CP asymmetries, namely,  
\begin{eqnarray}
	\label{isospin-sumrule}
	\nn && A_{\rm CP}(\pi^{-} K^{+})+A_{\rm CP}(\pi^{+} K^{0}) \frac{\mathcal{B}(\pi^{+} K^{0})\tau_{0}}
	{\mathcal{B}(\pi^{-} K^{+})\tau_{+}} = 
\\
&& \qquad A_{\rm CP}(\pi^{0} K^{+})\frac{2\mathcal{B}(\pi^{0} K^{+})\tau_{0}}{\mathcal{B}(\pi^{-} K^{+})\tau_{+}} + 
A_{\rm CP}(\pi^{0} K^{0})\frac{2\mathcal{B}(\pi^{0} K^{0})}{\mathcal{B}(\pi^{-} K^{+})}\,,
\end{eqnarray}
holds up to a few percent where $\mathcal{B}(\pi K)$ are the BRs and $\tau_{+}$ and $\tau_{0}$ are
the lifetimes of  $B^{+}$ and $B^{0}$ mesons respectively. While deriving Eq.~\eqref{isospin-sumrule}, 
it was assumed that the annihilation amplitude $A$ is suppressed relative to colour-allowed tree amplitude $T$ and 
the relative strong phase difference between the $T$ and $C$ amplitudes is small. The algebraic relation, 
Eq.~\eqref{T-PEW reln}, between tree and EWP amplitudes was also used. A deviation from 
the sum rule in Eq.~\eqref{isospin-sumrule} is quantified by an observable $\Delta_{4}$:
\begin{align}\label{eq:del4}
\nn \Delta_{4} &= A_{\rm CP}(\pi^{-} K^{+})+A_{\rm CP}(\pi^{+} K^{0})\frac{\mathcal{B}(\pi^{+} K^{0})\tau_{0}}{\mathcal{B}(\pi^{-} K^{+})\tau_{+}}  \\
& \qquad -A_{\rm CP}(\pi^{0} K^{+})\frac{2\mathcal{B}(\pi^{0} K^{+})\tau_{0}}{\mathcal{B}(\pi^{-} K^{+})\tau_{+}}-
A_{\rm CP}(\pi^{0} K^{0})\frac{2\mathcal{B}(\pi^{0} K^{0})}{\mathcal{B}(\pi^{-} K^{+})}\,,
\end{align}
so that $\Delta_4\not=0$ may be considered as a strong hint for NP. 
It has been shown in Ref.\ \cite{Imbeault:2006nx} that NP contributions in $B\to \pi K$ decays can
be absorbed by reparametrising the SM amplitudes, so that NP effects may be hidden. However, observables like $\delacp$ and $\Delta_4$ are not invariant under such reparametrisation.

\section{Methodology}
	\label{sec:methodology}
\begin{table}
	\centering
	\footnotesize
	\begin{ruledtabular}
		\begin{tabular}{c|rl|rl|c}	
			Modes& Expt.	& BR $[10^{-6}]$		& Expt. 	&$A_{\rm CP}$	&$S_{\rm CP}$\\ 
			\hline
			&	BaBar~\cite{Aubert:2006fha}			&	$19.1(6)(6)$		&	BaBar~\cite{Lees:2012mma}	&	$-0.107(16)(^{6}_{4})$	&	\\
			$B^{0}\to \pi^{-}K^{+}$		&	Belle~\cite{Duh:2012ie}		&	$20.00(34)(60)$		&	Belle~\cite{Duh:2012ie}	&	$-0.069(14)(7)$	&	\\
			& CLEO~\cite{Bornheim:2003bv}	&	$18.0(^{23}_{21}) (^{12}_{9})$	&	CDF~\cite{Aaltonen:2014vra}	&	$-0.083(13)(4)$	&	\\
			&	& &LHCb~\cite{Aaij:2018tfw}	&	$-0.084(4)(3)$	&	\\
			&	& &LHCb~\cite{Aaij:2020buf}	&	$-0.0824(33)(33)$	&	\\
			& Belle-II~\cite{Abudinen:2021gly}& $18.0(9)(9)$ &	Belle-II~\cite{Abudinen:2021gly}& $-0.16(5)(1)$ &\\		
			\hline
			&	BaBar~\cite{Aubert:2007hh}	&	$13.6(6)(7)$	&	BaBar~\cite{Aubert:2007hh}	&	$0.030(39)(10)$	&	\\
			$B^{+}\to \pi^{0}K^{+}$	&	Belle~\cite{Duh:2012ie}	&	$12.62(31)(56)$&	Belle~\cite{Duh:2012ie}	&			$0.043(24)(2)$	&	\\
			&	CLEO~\cite{Bornheim:2003bv}	&	$12.9(^{24}_{22})(^{12}_{11})$	&	LHCb~\cite{Aaij:2020wnj}	&	$0.025(15)(6)$	&	\\
			&	Belle-II~\cite{Abudinen:2021zmr}	&	$11.9(^{11}_{10})(^{16}_{11})$	&	Belle-II~\cite{Abudinen:2021zmr}	&	$-0.09(9)(3)$	&	\\
			\hline
			&	BaBar~\cite{Aubert:2006gm}	&	$23.9(11)(10)$	&	BaBar~\cite{Aubert:2006gm}	&	$-0.029(39)(10)$	&	\\
			$B^{+}\to \pi^{+}K^{0}$	&	Belle~\cite{Duh:2012ie}	&	$23.97(53)(71)$	&	Belle~\cite{Duh:2012ie}	&	$-0.011(21)(6)$	&	\\
			&	CLEO~\cite{Bornheim:2003bv}	&	$18.8(^{37}_{33})(^{21}_{18})$	&	LHCb~\cite{Aaij:2013fja}	&	$-0.022(25)(10)$	&	\\
			& Belle-II~\cite{Abudinen:2021gly}& $21.4(^{23}_{22})(16)$ &	Belle-II~\cite{Abudinen:2021gly}& $-0.01(8)(5)$ &\\				
			\hline
			&	BaBar~\cite{Lees:2012mma}	&	$10.1(6)(4)$	&	BaBar~\cite{Aubert:2008ad,HFLAV:2018corr}	&	$-0.13(13)(3)$	&	$0.55(20)(3)$~\cite{Aubert:2008ad,HFLAV:2018corr}	\\
			$B^{0}\to \pi^{0}K^{0}$	&	Belle~\cite{Fujikawa:2008pk}	&	$8.7(5)(6)$	&	Belle~\cite{Fujikawa:2008pk,HFLAV:2018corr}	&	$0.14 (13)(6)$	&	$0.67(31)(8)$~\cite{Fujikawa:2008pk,HFLAV:2018corr}	\\
			&	Belle~\cite{Duh:2012ie}	&	$9.68(46)(50)$	&	&	&	\\
			&	CLEO~\cite{Bornheim:2003bv}	&	$ 12.8(^{40}_{33})(^{17}_{14})$	&	&	&	\\
			&	Belle-II~\cite{Abudinen:2021dyj}	&	$8.5(^{17}_{16})(12)$	&	Belle-II~\cite{Abudinen:2021dyj}	&	$-0.40(^{46}_{44})(4)$	&	\\			
		\end{tabular}
	\end{ruledtabular}
	\caption{\small Experimental inputs used in this work. The first uncertainty is statistical and the second one, systematic.}
	\label{Tab:tab1}
\end{table}

	The goal of the numerical analysis would be to find out the posterior distribution (parameter space) of the various 
amplitudes and their corresponding relative phases, allowed by the data. The available data consist of 4 BRs for the
$B\to \pi K$ modes, 4 direct CP asymmetries ($A_{\rm CP}$) and the 
mixing-induced CP asymmetry $S_{\rm CP}$ measured for the $B\to \pi^{0}K^{0}$ decay. Instead of using the 
averages quoted in~\cite{Zyla:2020zbs}, we utilize all the available data that are used to calculate those averages 
for the fits and show them in Table~\ref{Tab:tab1}. Very recent results from Belle-II 
\cite{Abudinen:2021gly,Abudinen:2021zmr,Abudinen:2021dyj} are also included in this analysis. Without neglecting any 
contribution from diagrams up to $\mathcal{O}(\lambda^{3})$, 
we have 10 free parameters: the 5 magnitudes $P_{tc}$, $\vert T \vert,\,\vert C \vert,\,\vert A \vert,\, \vert P_{uc} \vert$, 
4 relative phases $\delta_{T}, \, \delta_{C}, \, \delta_{A},\, \delta_{P_{uc}}$ and the parameter $\kappa$, which we have 
chosen, at times, to vary as a free parameter, instead of fixing it to the SM expectation. We have defined the relative phases 
with respect to the $P_{tc}$ diagram whose absolute phase is set to zero in this convention~\cite{Kim:2007kx}. 

Let us explain the rationale for using all the data points for an observable, and not their quoted average. First, the
quoted averages do not include all the latest results from LHCb and Belle-II. Secondly, a correct averaging should 
implement all statistical and systematic correlations between the data-points, which we do not have at our disposal. 
Thus, we are forced to perform our own averaging, which involves 
the creation of a negative log-likelihood using all data points. This is precisely what we do in our analysis,
with the observables replaced by their parametric theoretical expressions. 

In the next Section, we talk about our fits. Among the 6 fits performed in this paper, only two {\em Order-3} fits have 
9 and 10 free parameters. For the rest, the number of free parameters is 6 or 7; an analysis with only the averages 
would have been perfectly possible for all of them, and even a frequentist fit would have been meaningful. 
Though a numerical minimization of a cost function 
(negative log-likelihood or $\chi^2$) can always be done with an arbitrary number of parameters and a best fit can be obtained, interpretation of that result (when the number of free parameters is more than the number of data points)
from a frequentist point of view is dicey, as effective degrees of freedom becomes unphysical. 
From a Bayesian point of view, however, this just means that the posteriors are unconstrained. We will show that this is 
exactly what we find for our {\em Order-3} fits, whose posteriors have more than optimal variance.

	Though we mainly follow a Bayesian framework for the purpose of the present analysis, we also 
simultaneously follow the frequentist interpretations of our results, whenever possible. This means that in addition to 
the obtained parameter-posteriors and the corresponding estimates of central tendency and dispersion, we also keep track 
of the maximum likelihood estimates (MLEs), corresponding fit-probabilities (in terms of $p$-values) and one-dimensional 
confidence levels (CLs). In the Appendix, we attach a glossary for the Bayesian terms used here.

Apart from the free parameters, the magnitude of the CKM elements 
$\vert V_{ub} \vert$, $\vert V_{us} \vert$, $\vert V_{tb} \vert$, 
$\vert V_{ts} \vert$, and the CKM angles $\beta$ and $\gamma$, come into the analysis as uncertain theoretical inputs 
coinciding with latest HFLAV averages~\cite{Amhis:2019ckw}. They are incorporated as multi-normal priors, details of 
which are provided in the lower part of the second column of Table \ref{Tab:tab2}. The effect of the uncertainties of 
meson masses and the $B$-meson lifetimes are very small compared to the other sources of uncertainties and are thus 
neglected in the present analysis.

	\paragraph{\underline{Details of Frequentist fit-procedure}:}
	Finding the MLE of the parameters boils down to minimising the quantity $\chi^2 \equiv - 2 \ln(\mathcal{L})$ with 
respect to the parameters, where $\mathcal{L}$ is the likelihood for the experimental observations. For Gaussian data, 
it simplifies to the actual form of a $\chi^2$ function. Except particular $S_{\rm CP}$ and $A_{\rm CP}$ measurements in BaBar 
and Belle which are correlated~\cite{HFLAV:2018corr}, the experimental inputs are independent and we incorporate 
that correlation in our analysis. 
Whenever applicable, we take average of the asymmetric uncertainties. 
One may note that the frequentist fits make sense only for the {\em Order-2} fits described in the next Section.

	\paragraph{\underline{Details of the Bayesian analysis}:}
	CKM parameters come into the analysis as theoretical inputs and as mentioned above, we use their measured values as 
multi-normal priors. For all other parameters, uniform priors are supplied in a wide range. Using the log-likelihood 
($\ln\mathcal{L}$) and the priors, we sample the un-normalised log-posterior by running a Markov Chain Monte Carlo 
(MCMC) process. We follow the Metropolis-Hastings algorithm \cite{MHAlgo} with a multi-normal proposal distribution 
for the MCMC runs. Convergence of the first quartile is ensured using single-chain diagnostics like Raftery-Lewis 
\cite{Raftery1991HowMI} and thinned samples are used to reduce the auto-correlation of the chain.

Our results on the allowed parameter spaces are shown, for clarity, as two-dimensional Bayesian fits. 
The fits have been organised on two factors: the relationship of the EWP amplitudes to $T$ and $C$ amplitudes, and 
the number of parameters considered for the fit. 
\begin{itemize}
\item The relationship shown in Eq.\ \ref{T-PEW reln} holds to a very good extent in the SM, leading to a single $\kappa$ 
parameter. It may not actually be so when NP is present, {\em e.g.}, there can be two such parameters, 
$P_{EW}\sim \kappa_1 T$ and $P_{EW}^{C}\sim \kappa_2 C$. This case, and the case with $\kappa$ as a complex 
parameter, have been considered in detail. Such options are, of course, not SM-like.
\item 
Among the 10 free parameters, two amplitudes, namely, $\vert A \vert$ and $\vert P_{uc} \vert$ are
expected to be small ($\sim \lambda^3$), and hence the exact values of their phases, $\delta_{A}$ and $\delta_{P_{uc}}$
respectively, are expected to be irrelevant. For all the cases of $\kappa$ discussed earlier, we have performed two sets of fits:
one with all the 9 parameters (plus $\kappa$), which we call {\em Order-3} fits, and one excluding the 
4 parameters mentioned above, which we call {\em Order-2} fits. 
\end{itemize}
The rationale for all these fits and the final results follow in Section \ref{sec:results}. By letting the parameters vary over 
a large range (not necessarily consistent with SM), we ensure to get the correct global picture of the potentially 
complicated multi-dimensional probability landscape of the parameters.

\section{Results}
	\label{sec:results}
\begin{figure}
	\centering
	\subfloat[$\kappa_1$ vs. $\kappa_2$]{\includegraphics[width=0.47\textwidth,height=0.22\textheight]{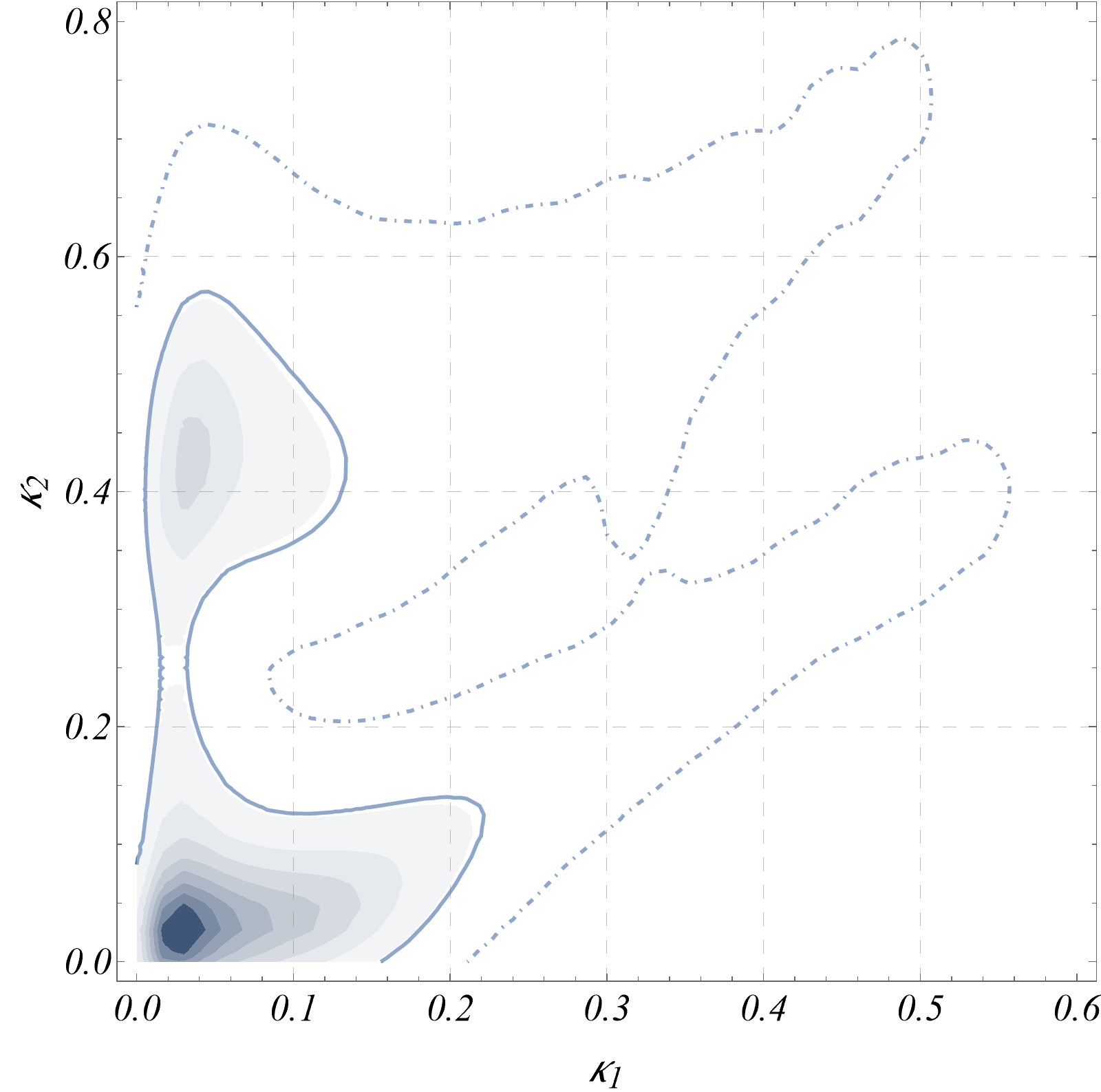}\label{fig:Alt2d11}}~~
	\subfloat[$\kappa_1$, $\kappa_2$ and $\kappa$ vs. $P_{tc}$]{\includegraphics[width=0.47\textwidth,height=0.22\textheight]{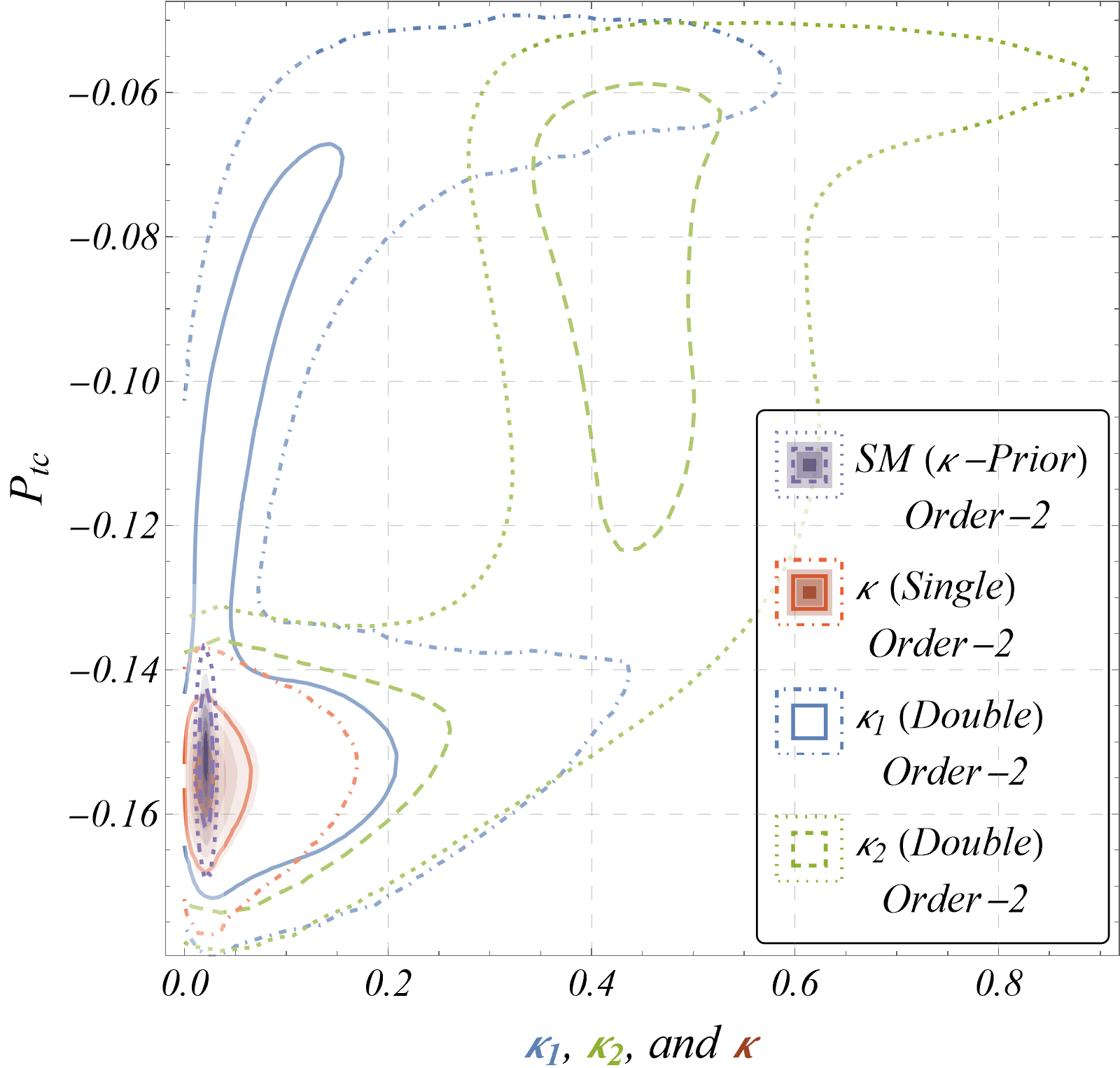}\label{fig:Alt2d12}}\\
	\subfloat[$\kappa_1$, $\kappa_2$ and $\kappa$ vs. $|T|$]{\includegraphics[width=0.47\textwidth,height=0.22\textheight]{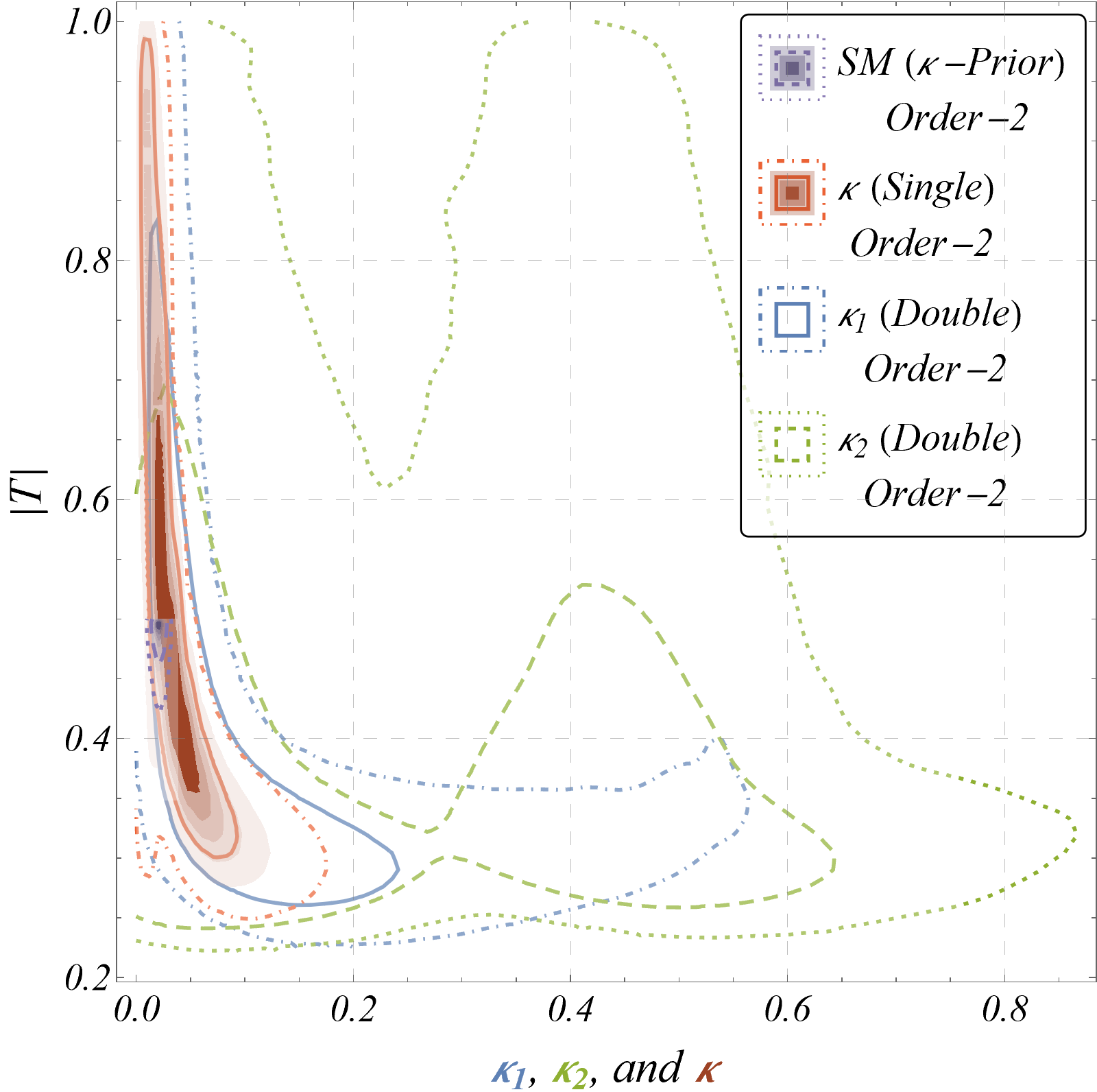}\label{fig:Alt2d13}}~~	
	\subfloat[$\kappa_1$, $\kappa_2$ and $\kappa$ vs. $|C|$]{\includegraphics[width=0.47\textwidth,height=0.22\textheight]{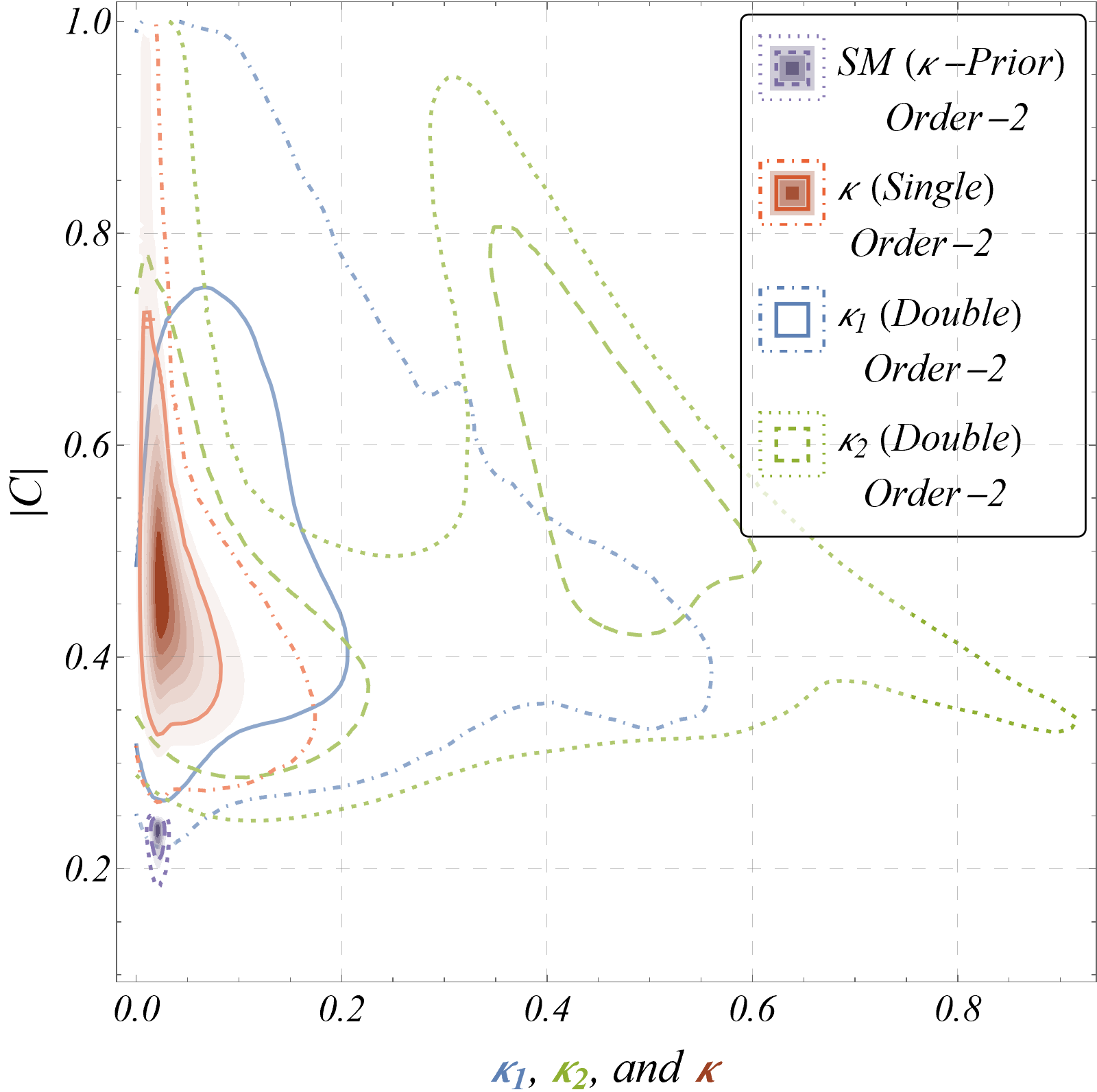}\label{fig:Alt2d14}}\\
	\subfloat[$\kappa_1$, $\kappa_2$ and $\kappa$ vs. $\delta_T$]{\includegraphics[width=0.47\textwidth,height=0.22\textheight]{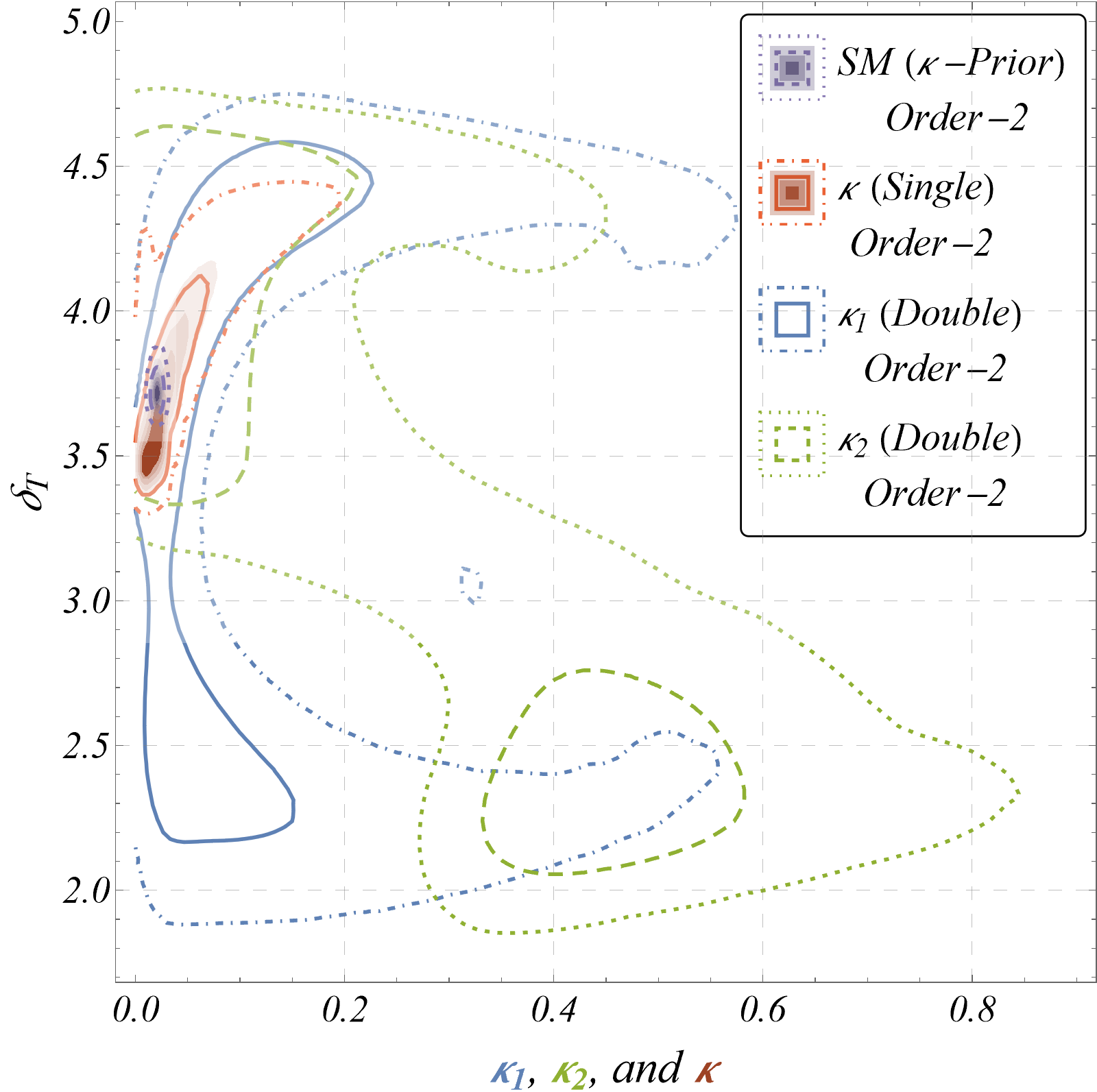}\label{fig:Alt2d15}}~~
	\subfloat[$\kappa_1$, $\kappa_2$ and $\kappa$ vs. $\delta_C$]{\includegraphics[width=0.47\textwidth,height=0.22\textheight]{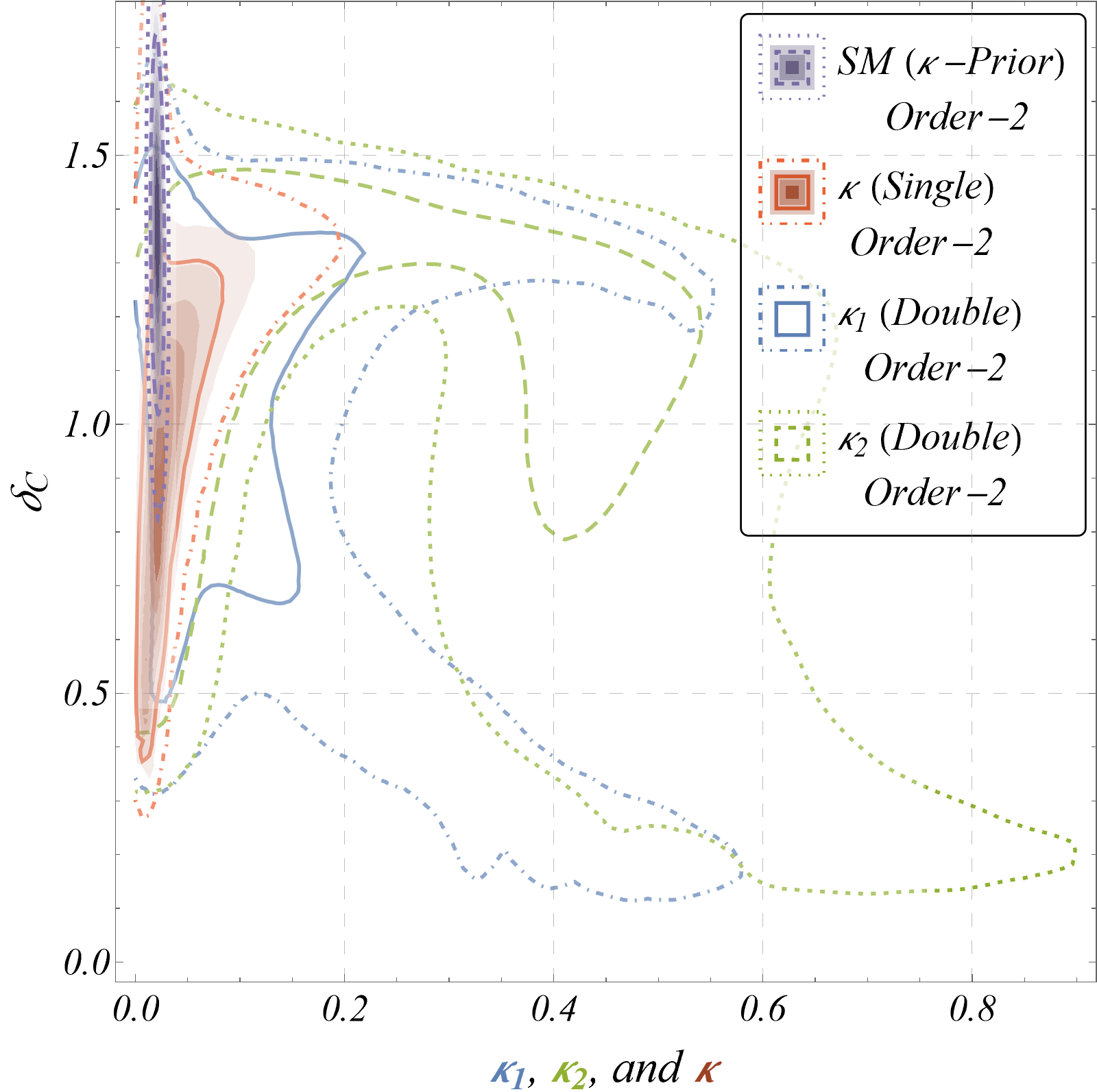}\label{fig:Alt2d16}}
	\caption{\footnotesize Marginal posteriors (2-D) with constant probability contours. In Fig.\ \ref{fig:Alt2d11}, contours enclose $68.28\%$ 
	(blue, solid) and $95.45\%$ (blue, dot-dashed) CIs. Those in rest of the figures are for $\kappa_1$ vs. the corresponding 
	parameter in the $y$-axis, while green ones, dashed and dotted, represent the $\kappa_2$ marginals. The 
	reddish brown contours with changing opacity enclose regions with decreasing probability (from darker to lighter) 
	for the {\em Order-2} fit with a single real $\kappa$.}
	\label{fig:Alt2d1}
\end{figure}

\begin{figure}
	\centering
	\subfloat[$P_{tc}$ vs. $|T|$]{\includegraphics[width=0.47\textwidth,height=0.23\textheight]{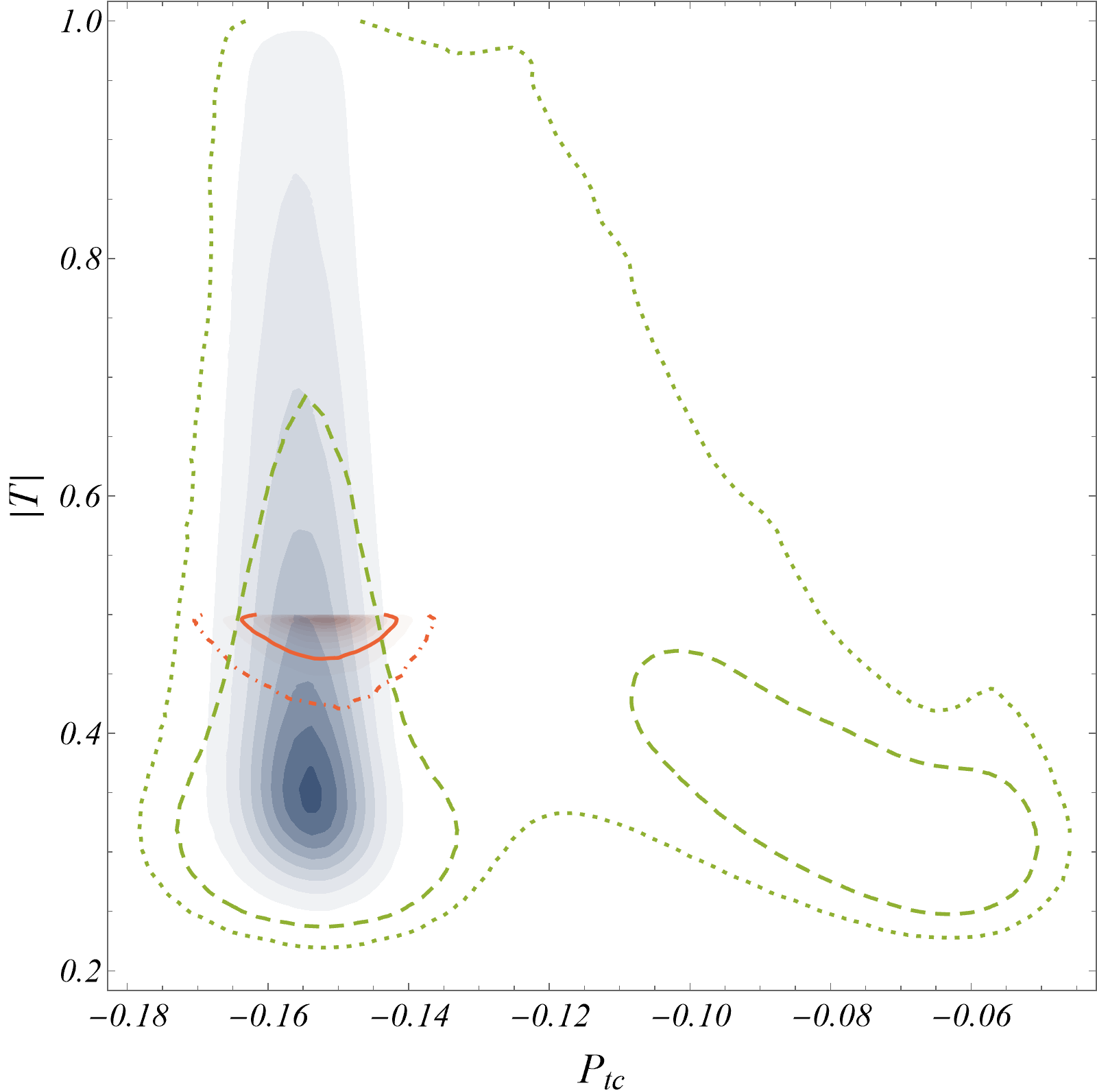}\label{fig:Alt2d23}}~~
	\subfloat[$P_{tc}$ vs. $|C|$]{\includegraphics[width=0.47\textwidth,height=0.23\textheight]{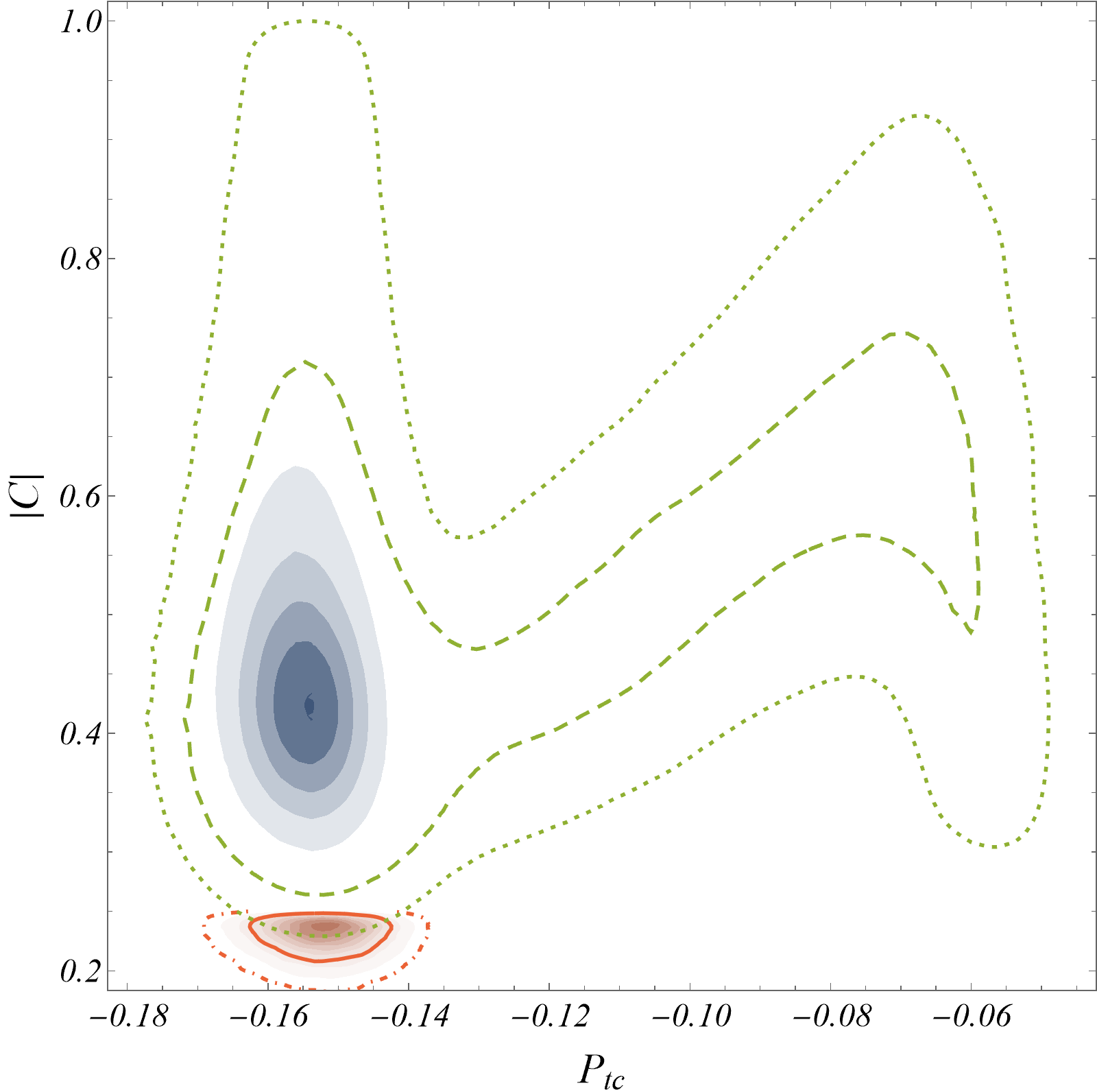}\label{fig:Alt2d24}}\\
	\subfloat[$P_{tc}$ vs. $\delta_T$]{\includegraphics[width=0.47\textwidth,height=0.23\textheight]{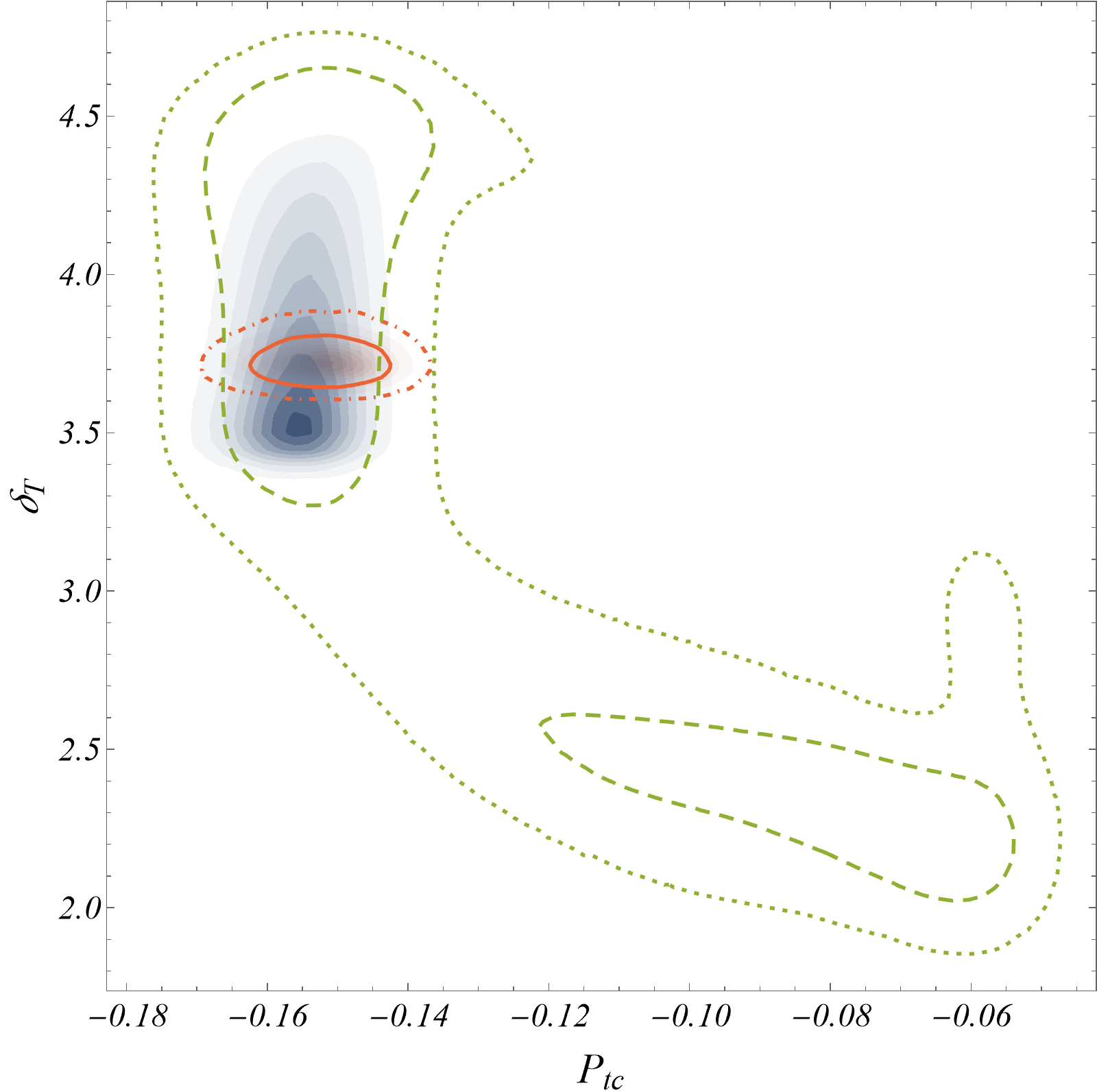}\label{fig:Alt2d25}}~~
	\subfloat[$P_{tc}$ vs. $\delta_C$]{\includegraphics[width=0.47\textwidth,height=0.23\textheight]{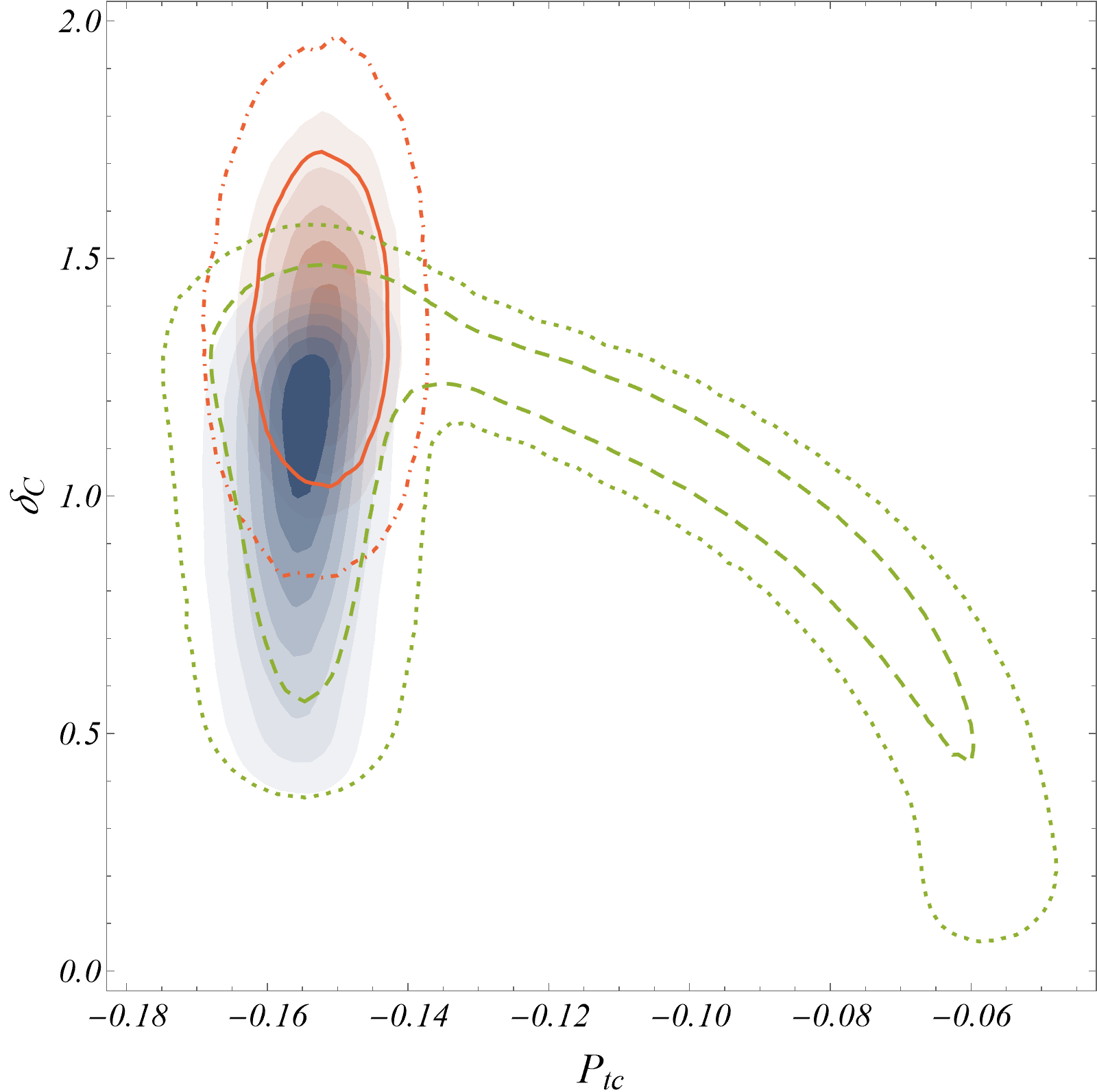}\label{fig:Alt2d26}}\\
	\subfloat[$|T|$ vs. $|C|$]{\includegraphics[width=0.47\textwidth,height=0.23\textheight]{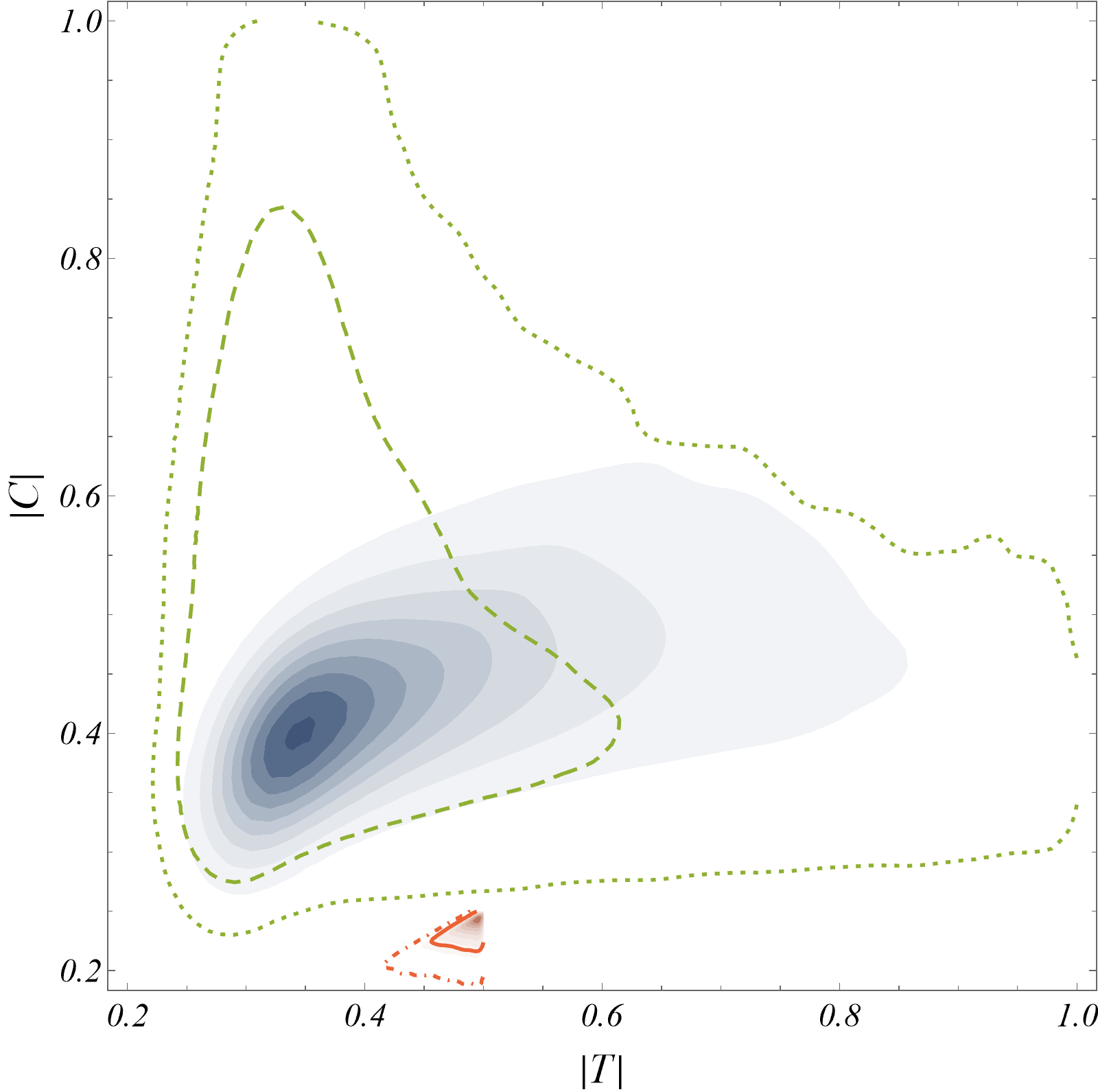}\label{fig:Alt2d34}}~~~~~~
	\subfloat[Legend]{\includegraphics[height=0.17\textheight]{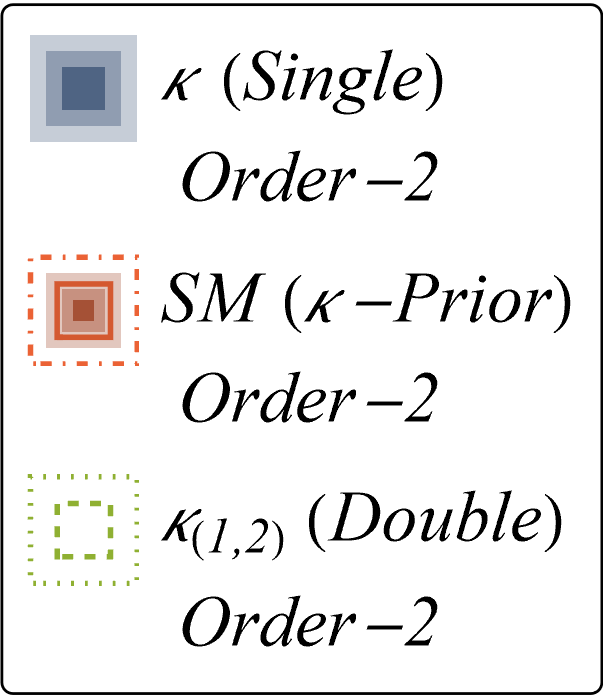}\label{fig:Alt2dleg1}}
	\caption{\footnotesize Marginal posterior distributions (2-D) with constant probability contours. Green dashed (dotted) contours 
	represent the $68.28\%$ ($95.45\%$) CIs for the \textit{Double $\kappa$ Order-2} fit, while red (solid, dot-dashed) contours 
	denote those of the \textit{$\kappa$-Prior Order-2} fit. The bluish contours with changing opacity enclose the high 
	probability regions with decreasing probability (from darker to lighter) for the \textit{Single real $\kappa$ Order-2} fit.}
	\label{fig:Alt2d2}
\end{figure}

\begin{figure}
	\centering
	\subfloat[$|T|$ vs. $\delta_T$]{\includegraphics[width=0.45\textwidth,height=0.25\textheight]{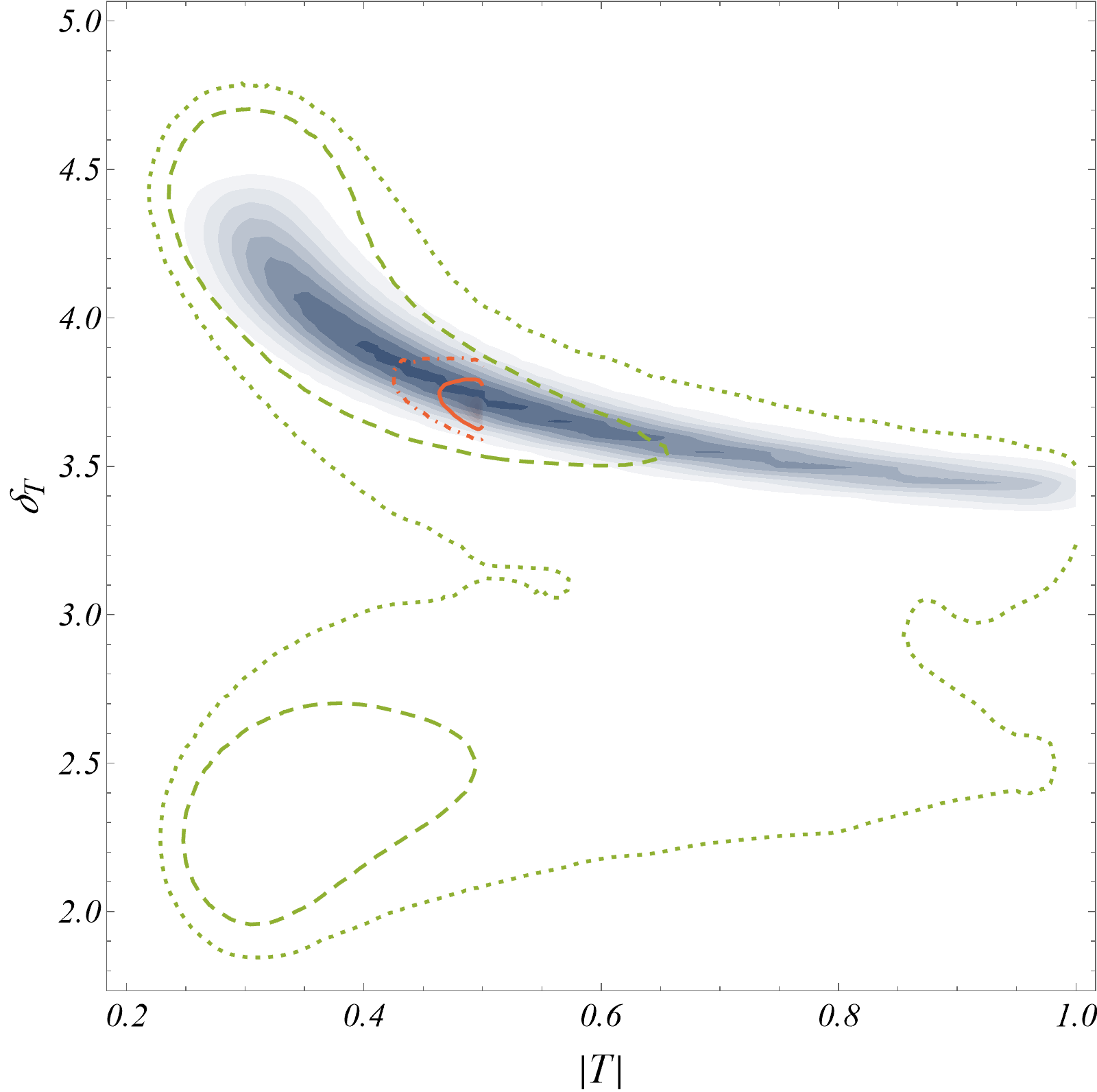}\label{fig:Alt2d35}}~~~~
	\subfloat[$|T|$ vs. $\delta_C$]{\includegraphics[width=0.45\textwidth,height=0.25\textheight]{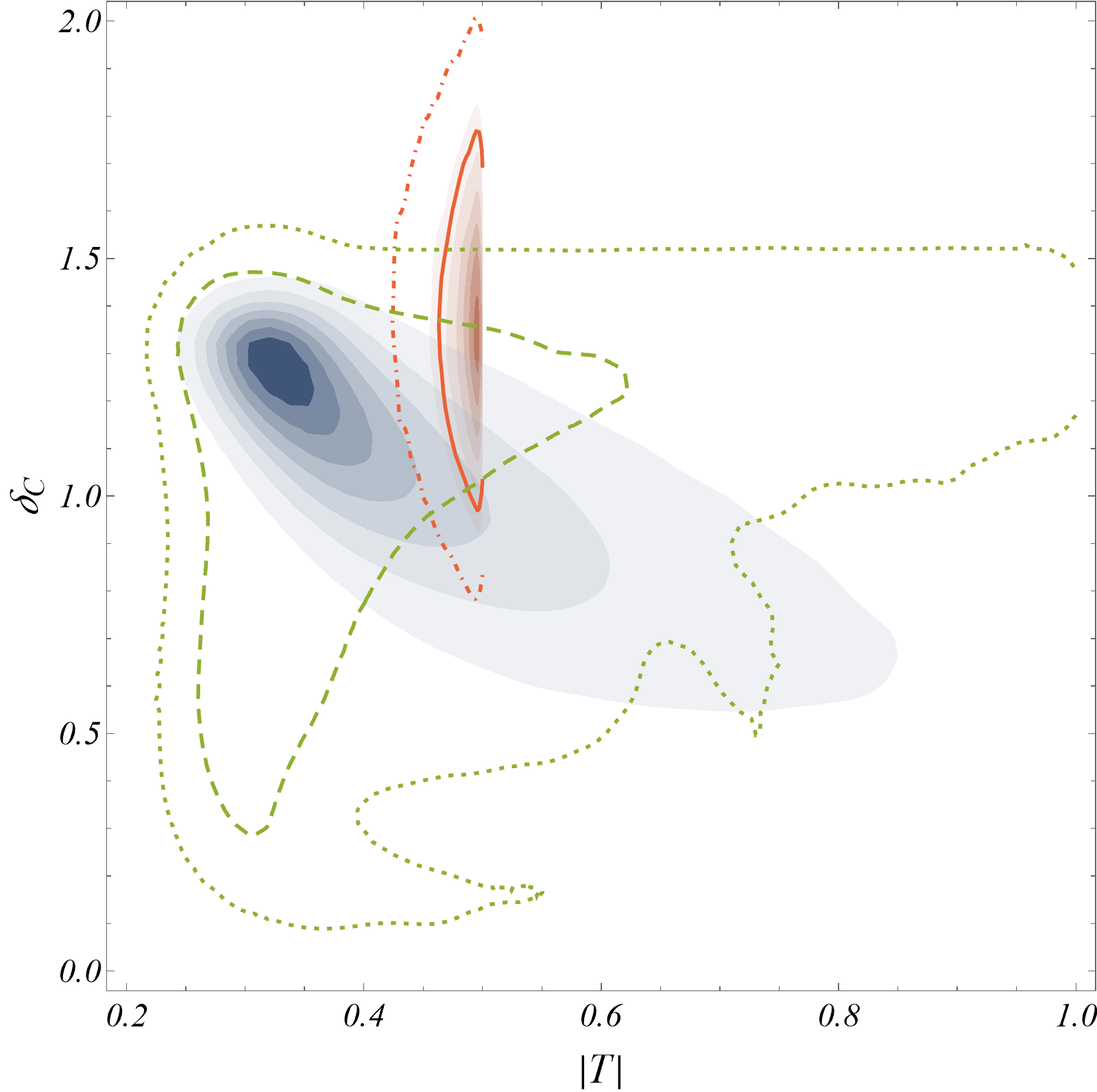}\label{fig:Alt2d36}}\\
	\subfloat[$|C|$ vs. $\delta_T$]{\includegraphics[width=0.45\textwidth,height=0.25\textheight]{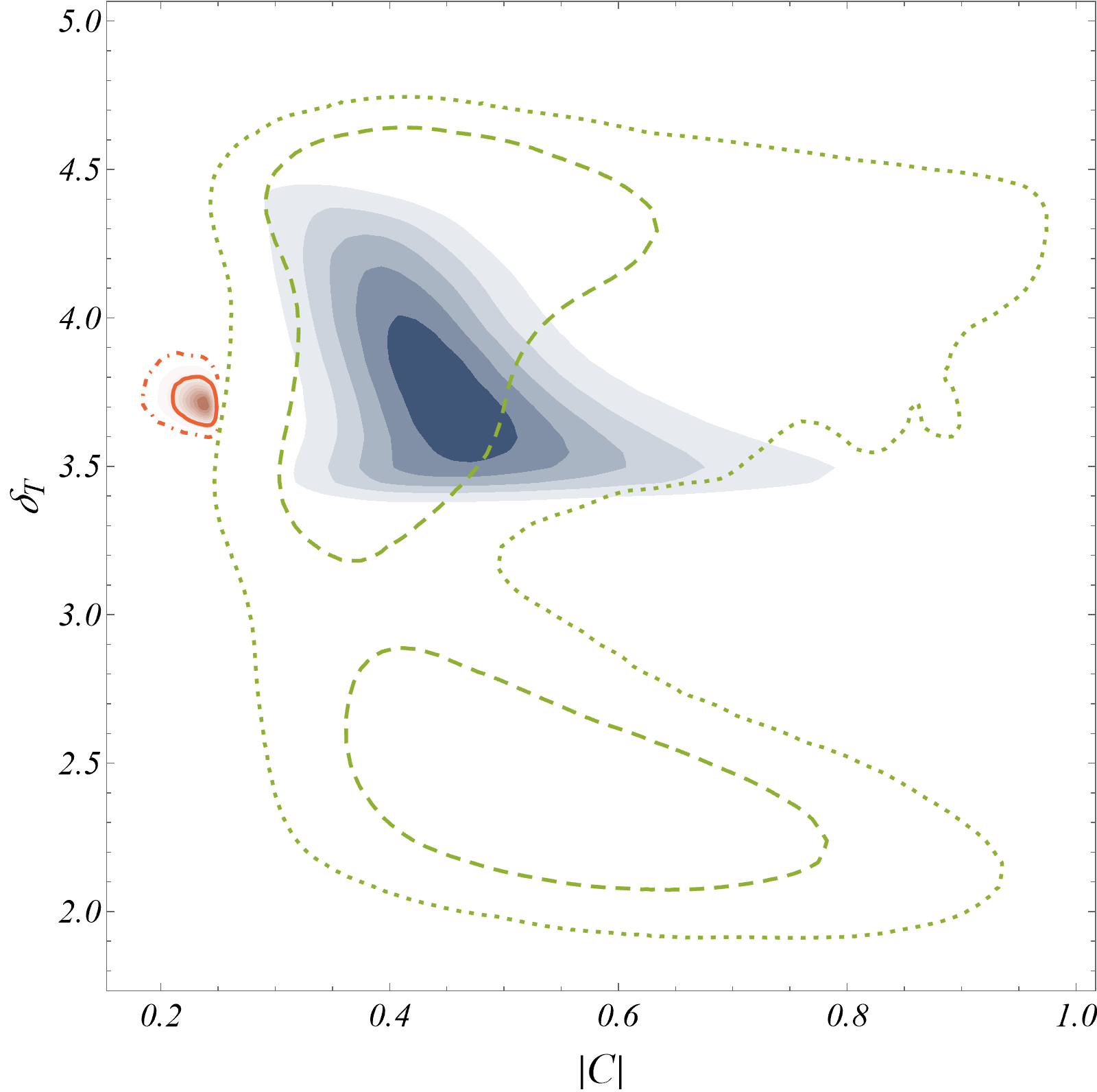}\label{fig:Alt2d45}}~~~~
	\subfloat[$|C|$ vs. $\delta_C$]{\includegraphics[width=0.45\textwidth,height=0.25\textheight]{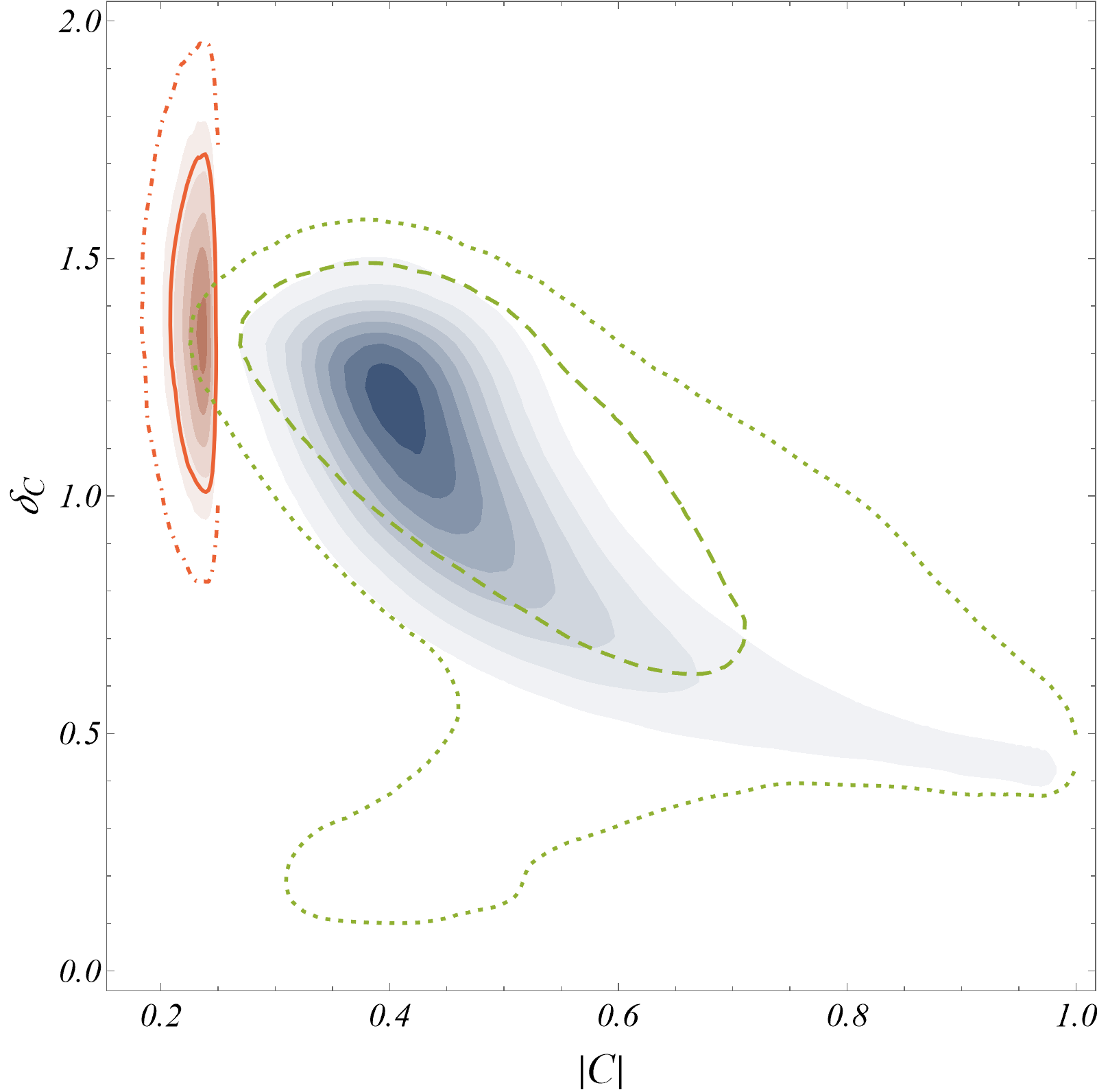}\label{fig:Alt2d46}}\\
	\subfloat[$\delta_T$ vs. $\delta_C$]{\includegraphics[width=0.45\textwidth,height=0.25\textheight]{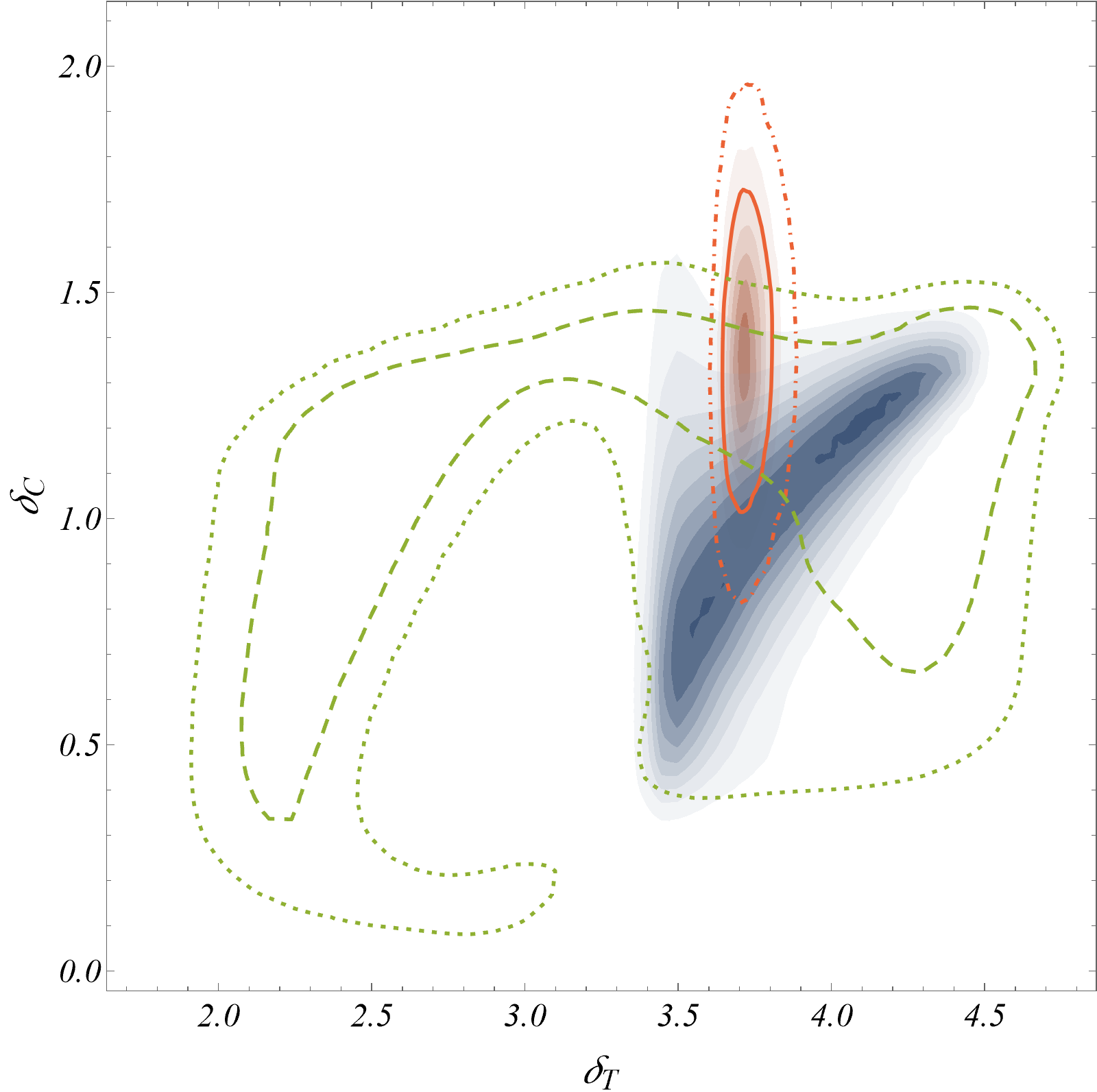}\label{fig:Alt2d56}}~~~~~~
	\subfloat[Legend]{\includegraphics[height=0.17\textheight]{fig3f}\label{fig:Alt2dleg}}
	\caption{\footnotesize Continued from Fig.\ \ref{fig:Alt2d2}.}
	\label{fig:Alt2d3}
\end{figure}

We will show our results for 6 different fits, as follows:\\
(i) {\em Order-2}, with a single real $\kappa$ as a free parameter;\\
(ii) Same as (i) but at {\em Order-3};\\
(iii) {\em Order-2}, with a single real $\kappa$ as a given normal prior ($0.014\pm 0.006$), and with the extra constraint 
$|C|\leq |T|/2$;\\
(iv) Same as (iii) but at {\em Order-3};\\
(v) {\em Order-2}, with a single complex $\kappa = |\kappa|\exp(i\delta_\kappa)$ as a free parameter;\\
(vi) {\em Order-2}, with two real $\kappa$-type free parameters, $\kappa_1$ and $\kappa_2$. 

Let us first summaries the main points of this Section:\\
1. The {\em Order-2} fit with a single real $\kappa$ as a given prior produces a perfectly acceptable fit in the 
SM-like region. When $\kappa$ is treated as a free parameter, the best-fit region is found to have
a large overlap with the SM-like parameter space. \\
2. The {\em Order-3} fits do not improve over the {\em Order-2} fits. In other words, {\em Order-2} 
seems to be the optimal choice at the present experimental precision.\\
3. The predictions for $\delacp$ and $\Delta_4$ are mutually consistent for all the fits.\\
4. The fit with two real $\kappa$ parameters  shows some interesting features while keeping the 
above conclusions more or less intact, but this definitely falls under NP.

We will now quantify these qualitative remarks. 

\subsection{Order-2 fits}\label{sec:res_paramsp}

\begin{table}
	\centering
	\begin{ruledtabular}
		\renewcommand*{\arraystretch}{1.15}
		\begin{tabular}{c c c c c c c}
			Parameters & Priors & \multicolumn{4}{c|}{Real $\kappa$} & Complex $\kappa$ \\
			\cline{3-7}
			& & \multicolumn{2}{c|}{SM ($\kappa$ Prior)} & \multicolumn{3}{c}{$\kappa$ Free} \\
			\cline{3-7}
			& & Order-2 & Order-3 & Order-2 & Order-3 & Order-2 \\
			\hline
			$\kappa$  &  $0.014(6)$  &   $0.0210(_{43}^{44})$  &  $0.0210(_{43}^{44})$  &  $0.028(_{14}^{41})$  &  $0.029(_{14}^{47})$  &  $0.048(_{28}^{80})$  \\
			$P_{\text{tc}}$  &  -  &  $-0.1524(_{65}^{62})$  &  $-0.1524(_{66}^{61})$  &  $-0.1551(_{69}^{66})$  &  $-0.1548(_{70}^{72})$  &  $-0.1534(_{74}^{78})$  \\
			$\text{$|$T$|$}$  &  -  &  $0.486(_{22}^{11})$  &  $0.486(_{22}^{11})$  &  $0.49(_{15}^{28})$  &  $0.49(_{16}^{29})$  &  $0.68(_{24}^{22})$  \\
			$\text{$|$C$|$}$  &  -  &  $0.23(_{18}^{11})$  &  $0.23(_{18}^{12})$  &  $0.454(_{83}^{150})$  &  $0.471(_{94}^{166})$  &  $0.58(_{16}^{22})$  \\
			$\delta _{\kappa }$  &  -  &  -  &  -  &  -  &  -  &  $0.70(_{50}^{71})$  \\
			$\text{$|$A$|$}$  &  -  &  -  &  $0.0051(_{35}^{34})$  &  -  &  $0.047(_{32}^{35})$  &  -  \\
			$\left|P_{\text{uc}}\right|$  &  -  &  -  &  $0.0050(34)$  &  -  &  $0.049(_{33}^{35})$  &  -  \\
			$\delta _T$  &  -  &  $3.724(_{49}^{59})$  &  $3.724(_{51}^{59})$  &  $3.75(_{24}^{37})$  &  $3.71(_{23}^{37})$  &  $3.48(_{40}^{29})$  \\
			$\delta _C$  &  -  &  $1.37(_{22}^{23})$  &  $1.36(_{22}^{23})$  &  $1.03(_{34}^{26})$  &  $1.02(_{37}^{26})$  &  $0.74(_{33}^{540})$  \\
			$\delta _A$  &  -  &  -  &  $2.7(_{18}^{25})$  &  -  &  $3.7(_{26}^{17})$  &  -  \\
			$\delta _{P_{\text{uc}}}$  &  -  &  -  &  $3.3(_{23}^{20})$  &  -  &  $3.9(_{24}^{16})$  &  -  \\
			\hline
			$\left|V_{\text{us}}\right|$  &  $0.2245(8)$  &  $0.22455(_{81}^{80})$  &  $0.22456(80)$  &  $0.2245(8)$  &  $0.22449(80)$  &  $0.22451(80)$  \\
			$\left|V_{\text{ub}}\right|$  &  $0.0038(2)$  &  $0.00407(23)$  &  $0.00407(23)$  &  $0.00382(24)$  &  $0.00381(24)$  &  $0.00383(24)$  \\
			$\left|V_{\text{tb}}\right|$  &  $1.01(3)$  &  $1.016(30)$  &  $1.015(30)$  &  $1.011(30)$  &  $1.011(30)$  &  $1.011(30)$  \\
			$\left|V_{\text{ts}}\right|$  &  $0.039(1)$  &  $0.0389(11)$  &  $0.0389(11)$  &  $0.0387(11)$  &  $0.0387(11)$  &  $0.0387(11)$  \\
			$\gamma$  &  $1.26(7)(8)$  &  $1.247(91)$  &  $1.246(_{91}^{92})$  &  $1.24(11)$  &  $1.24(11)$  &  $1.24(11)$  \\
			$\beta$  &  $0.39(1)(1)$  &  $0.385(17)$  &  $0.385(17)$  &  $0.386(17)$  &  $0.386(17)$  &  $0.386(17)$  \\
		\end{tabular}
	\end{ruledtabular}
	\caption{\small Central tendency (Median) and uncertainties ($1\sigma\equiv 68.27\%$ Credible Intervals (CI) around the central estimates) for fits with one distinct high-probability region. Notably, uncertainty propagation using these parameters uses the whole sample from the posterior, not these point estimates. Parameters in the lower part of the table are used as priors in the fits. The topological amplitudes are defined by factoring out $G_{F}/\sqrt{2}$ and therefore the amplitudes are given in units of $(\text{GeV})^{3}$. }
	\label{Tab:tab2}
\end{table}

Let us start with fit (i), {\em i.e.}, the {\em Order-2} fit with three independent amplitudes, 
two phases, and $\kappa$. Just to cross-check with the frequentist approach, we find 
that the high-probability region of the parameter posterior distribution of the Bayesian analysis
 also contains the frequentist best-fit corresponding to a very high $p$-value.

The point estimates of the central tendency and dispersion of the parameters (medians with $1\sigma$ credible intervals
(CI) around them) are listed in the fifth column of Table \ref{Tab:tab2} (Real $\kappa$ (Free) Order-2). The 2D 
marginal-posteriors of the parameters are shown in Figs.\ \ref{fig:Alt2d12}--\ref{fig:Alt2d16} as reddish constant probability 
contours with increasing probability content (from lighter to darker), and as their bluish versions in Figs.\ \ref{fig:Alt2d2} and \ref{fig:Alt2d3}. 

A rather naive {\em Order-2} SM fit was performed first, setting the free parameters to vary within the following bounds:
\begin{align}
	0\leq\kappa\leq 0.03\,,\ \  -0.3\leq P_{tc}\leq 0\,,\ \  0\leq|T|\leq 0.5\,,\ \  0\leq|C|\leq 0.1\,.
\end{align}
We find that there is no acceptable fit in this region ($p$-value $< 1\%$). This holds true even if we extend the analysis to 
\textit{Order-3} and scan the 4 additional parameters in the following way:
\beq
0\leq|A|\leq 0.01\,,\ \  0\leq|P_{uc}|\leq 0.01\,,\ \  0\leq\delta_A\leq 2\pi\,,\ \  0\leq\delta_{P_{uc}}\leq 2\pi\,.
\label{eq:param-ord3}
\eeq
In other words, such a naive-SM parameter space is definitely ruled out by the data. 
	
Next, we focus on $\kappa$ and $C$, and take some leeway from the naive estimates.
Using the approximate value of $\kappa$ mentioned in Section \ref{sec:groundwork}, we use a 
normal prior for $\kappa$ in the fit, with the same median, but 5 times the uncertainty [fit (iii)]. This prior is 
mentioned in the second column of Table \ref{Tab:tab2}. Throughout this fit, we use the following additional constraints:
\beq 
-0.3\leq P_{tc}\leq 0\,, \ \ \  0\leq|T|\leq 0.5\,, \ \ \ |C|\leq (|T|/2)\,,
\eeq
{\em i.e.}, the tight $|C|<0.1$ being relaxed without jeopardising the SM expectation for the ratio.
It can, therefore, be argued, that any allowed parameter space we find in this region, may safely 
be labelled SM-like. Let us call this the {\em SM-like $\kappa$-prior Order-2} fit.
In a frequentist fit, this corresponds to a quite acceptable $p$-value.
The medians of the 1D-marginals 
with $1\sigma$ CIs around them are listed in the third column of Table \ref{Tab:tab2}. The constant (red) probability contours enclosing 
respectively the $68.28\%$ (solid) and $95.45\%$ (dot-dashed) CIs are shown as 2D marginal posteriors 
(with contours enclosing gradually increasing total probability-content from darker to lighter) in all the plots of Figs.\ \ref{fig:Alt2d2} and \ref{fig:Alt2d3}. As expected, this parameter
space has some overlap with that of the real $\kappa$ \textit{Order-2} fit. 
Similar 2-D marginals containing the parameter $\kappa$ are shown with similar purple contours in figure \ref{fig:Alt2d1}.

\begin{figure}
	\centering
	\subfloat[$|A|$ vs. $|P_{uc}|$]{\includegraphics[width=0.45\textwidth,height=0.3\textheight]{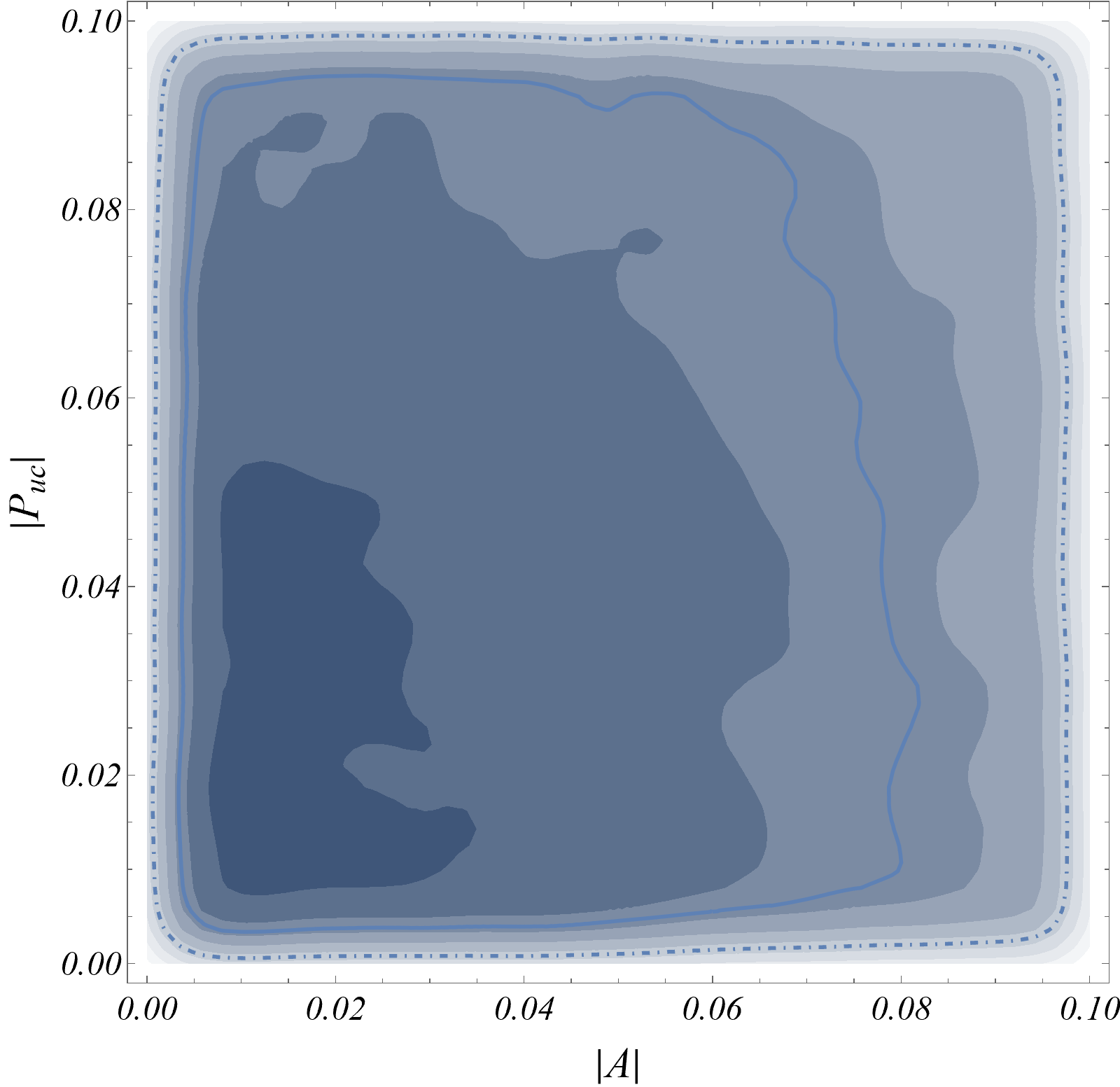}\label{fig:main2d56}}~~
	\subfloat[$\delta_A$ vs. $\delta_{P_{uc}}$]{\includegraphics[width=0.45\textwidth,height=0.3\textheight]{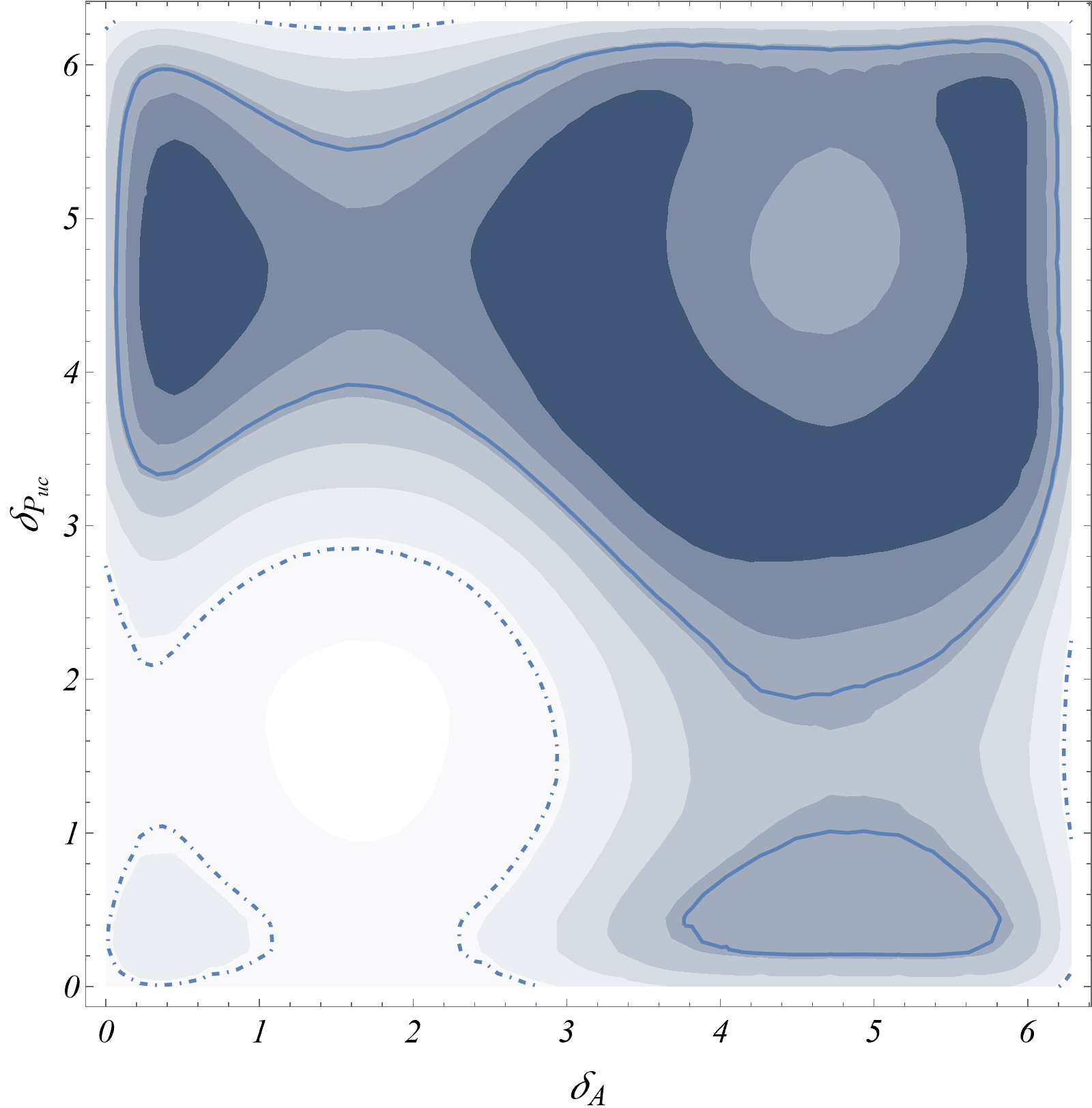}\label{fig:main2d910}}\\	
	\subfloat[$|\kappa|$ vs. $\delta_{\kappa}$]{\includegraphics[width=0.45\textwidth,height=0.3\textheight]{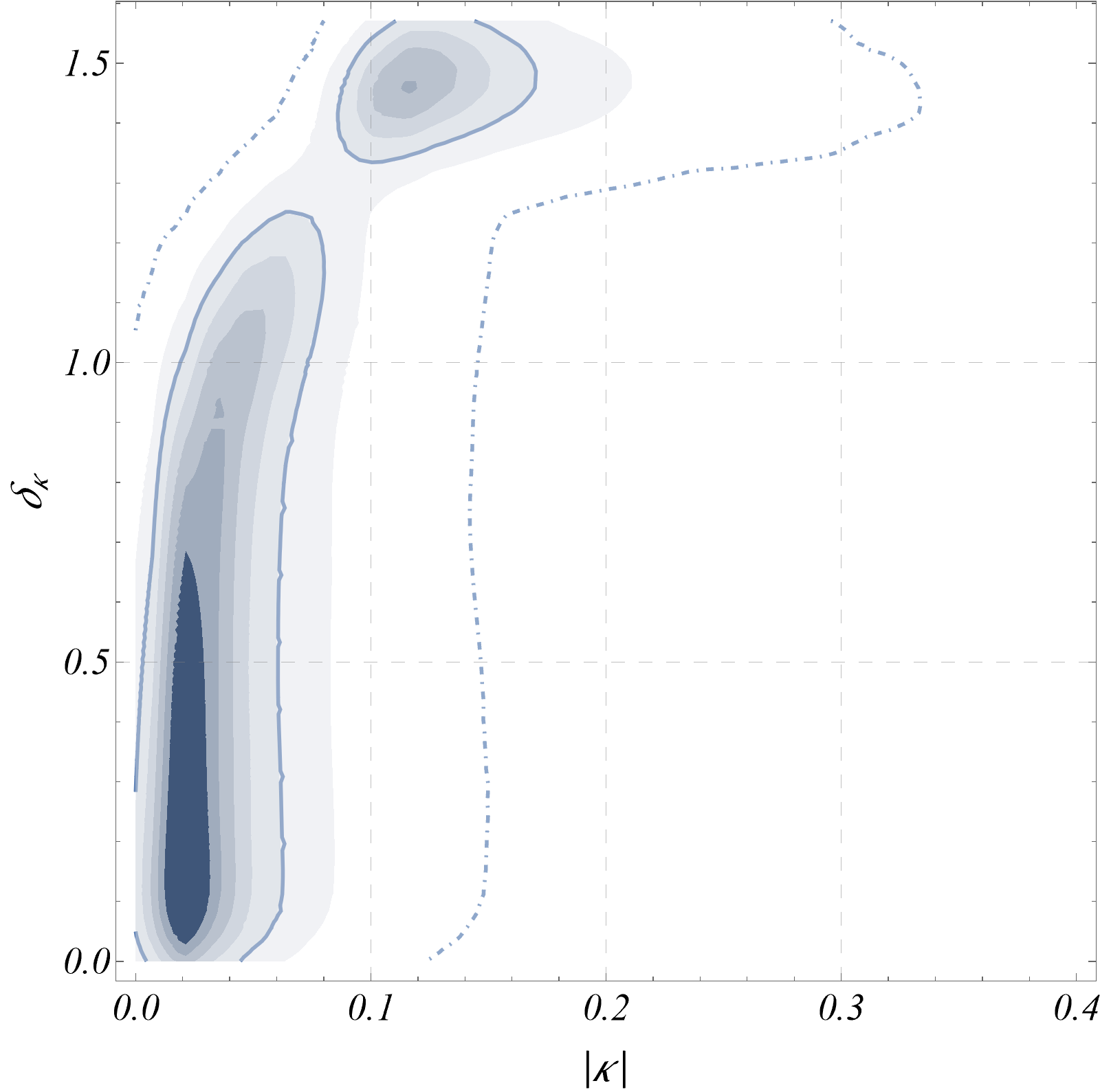}\label{fig:complex15}}
	\caption{\footnotesize 2D marginal posteriors similar to figure \ref{fig:Alt2d1}. First two figures are for the extra 4 parameters for the \textit{Single Real $\kappa$ Order-3 Fit}. The last one is the $|\kappa|$--$\delta_{\kappa}$ plane for the \textit{Complex $\kappa$ Order-2 Fit}.}
	\label{fig:Others2d1}
\end{figure}

In the next stage, we consider fit (vi), {\em i.e.}, two parameters $\kappa_1$ and $\kappa_2$ instead of a single $\kappa$
and perform the \textit{Order-2} fit. From the frequentist result, we find two distinct minima here (one deeper than the other): 
one where $\kappa_1\sim \kappa_2$ and another where $\kappa_2$ is considerably larger than $\kappa_1$. For both best-fits, 
$\kappa_1$-values are similar, and both are excellent fits. As can be seen from Figs.\ \ref{fig:Alt2d11} 
and \ref{fig:Alt2d12}, the Bayesian posteriors (in Fig.\ \ref{fig:Alt2d11}, blue solid ($68.28\%$) and dot-dashed ($95.45\%$) 
contours; in Figs.\ \ref{fig:Alt2d12}--\ref{fig:Alt2d16}, same for $\kappa_1$ and green dashed ($68.28\%$) and dotted ($95.45\%$) 
contours for $\kappa_2$) are consistent with this finding. 

Let us note here the ranges over which fit-parameters other than $\kappa$ are scanned, which clearly highlight the 
beyond-SM nature of the parameter space: 
\beq
-0.3\leq P_{tc}\leq 0\,,\ \  0\leq|T|\leq 1\,, \ \ 0\leq|C|\leq 1\,,\ \  0\leq\delta_T\leq 2\pi\,,\ \  0\leq\delta_C\leq 2\pi\,. 
\eeq
From figures \ref{fig:Alt2d1} and \ref{fig:Alt2d2}, one may see that there are two distinct high probability regions, 
separated at $1\sigma$ but connected at higher $\sigma$s in the $P_{\rm tc}$ direction, in the parameter space. 
The actual best fit in Bayesian approach has considerable overlap with the SM-like region of the single real $\kappa$ fit, 
except for the parameter $|C|$. This is evident from Figs.\ \ref{fig:Alt2d14}, \ref{fig:Alt2d24}, \ref{fig:Alt2d34}, 
\ref{fig:Alt2d45}, and \ref{fig:Alt2d46}, where the agreement is seen to be at $\sim 2\sigma$. 
One may note the physics behind this. The SM-like parameter-space is actually very skewed, due to the abrupt cut that we had imposed on $|C|$ and $|T|$ ($|T|\le 0.5$ and $|C|\le (|T|/2)$). Relaxing both these constraints even slightly should
naturally include a higher-probability parameter space. Following the discussion in Section \ref{sec:groundwork},
we see that such possibilities may not be entirely ruled out, in which case these regions, can, in turn, be 
completely consistent with those for the free real $\kappa$ case.
The other high probability region (containing the actual maximum likelihood estimate from the frequentist fit) is clearly 
far away from the SM expectations for the amplitudes. Where $\kappa_1$ shares most of its high-probability parameter
space with that of the single real $\kappa$, a set of quite large values of $\kappa_2$ are allowed in the non-standard 
region. A significant variation of $\kappa$ from Eq.\ (\ref{eq:kappa-def}) indicates that 
the WCs for the EWP operators are enhanced from their SM values, most probably because of NP.

If we vary $P_{tc}$ to include positive values as well, we will get a disjointed but identical parameter space 
symmetric about $P_{tc} = 0$. Since the strong phase of $P_{tc}$ is set to zero and the rest of the phases 
are defined relative to it, this sign ambiguity should be resolved by keeping only the positive or negative set 
of values of $P_{tc}$. We chose the negative part of the parameter space arbitrarily.

We can go further, consider fit (v), {\em i.e.}, 
take $\kappa$ to be a complex parameter $\kappa = |\kappa| \exp(i \delta_{\kappa})$
for the sake of the analysis, and repeat fit (i). The constant (blue) probability contours enclosing respectively the $68.28\%$ (solid) and $95.45\%$ (dot-dashed) credible regions in 2D marginal posteriors in the $|\kappa|$--$\delta_{\kappa}$ plane are shown in Fig.\ \ref{fig:complex15}, where $\delta_{\kappa}$ is equivalent to the relative strong phase between tree and electroweak penguin amplitudes, and is expected to be close to zero~\cite{Neubert:1998pt,Neubert:1998re}. Due to this, unlike the other phases occurring in this analysis, the prior for $\delta_{\kappa}$ is set to be a uniform distribution $0\leq\delta_{\kappa}\leq\frac{\pi}{2}$. The central tendency and dispersion of the posterior parameter space is depicted in the last column of Table \ref{Tab:tab2} in terms of the median and $1\sigma$ CIs around it. It is evident that the best fit solution is consistent with $\delta_\kappa=0$.

\subsection{Order-3 fits}\label{sec:res_paramsp3}

We can perform an analogous exercise, {\em i.e.}, fits (ii) and (iv), taking into account the four parameters neglected so far, 
and scanning them over the range
\beq
0\leq|A|\leq 0.1\,,\ \  0\leq|P_{uc}|\leq 0.1\,,\ \  0\leq\delta_A\leq 2\pi\,,\ \  0\leq\delta_{P_{uc}}\leq 2\pi\,.
\eeq
As we have pointed out earlier, a frequentist {\em Order-3} fit is questionable, and a Bayesian fit leads to 
unconstrained posteriors.

In other words, we expect the newly introduced $\mathcal{O}(\lambda^3)$ parameters to be very imprecise 
and the amplitudes to be consistent with zero. This is supported from the corresponding entries of the sixth column 
of Table \ref{Tab:tab2} (Real $\kappa$ (Free) Order-3) as well as from Figs.\ \ref{fig:main2d56} and 
\ref{fig:main2d910}, which show the 2D marginal-posteriors of the higher-order parameters as reddish constant 
probability contours. Though apparently the highest probability regions are close to zero, note that we have only 
scanned these amplitudes up to $0.1$, and observe a more or less flat nature of the 1D-posterior in that region. 
We refrain from showing the posteriors for the other parameters, as they are very similar to those of 
\textit{Order-2} fits, with only a slight increase in spread for some parameters.

The posterior medians with $1\sigma$ CIs for fit (iv), \textit{SM-like ($\kappa$-prior) Order-3}, are listed in the fourth 
column of Table \ref{Tab:tab2}. The {\em Order-2} fits being quite optimal at the present level, we have not tried 
the {\em Order-3} analogues for fits (v) and (vi), {\em i.e.}, with a complex $\kappa$ and with two real $\kappa$s. 

At this point, a few comments on the fit results are in order. 
From the unconstrained fit, we see that for the best-fit region, the colour-suppressed tree amplitude $|C|$ 
 is enhanced and is of the same order of magnitude as the colour-allowed tree amplitude $|T|$, resulting in 
$0.7 < |C|/|T| < 1$ for all the fits in Table \ref{Tab:tab2}, except the SM-like fits, where we had specifically 
set $|C|/|T| < 0.5$. This is consistent with the predictions from specific model
calculations~\cite{Bauer:2005kd,Huitu:2009st} for the topological amplitudes. On the other hand, where Soft Collinear 
Effective Theory based approaches predict $Arg(C/T) \sim 0$~\cite{Bauer:2005kd}, a quite large value of 
$Arg(C/T)$ ($\sim 200\degree$) is found to be favoured in the previous $B\to \pi K$ fits~\cite{Beaudry:2017gtw}. 
Non-zero values of $Arg(C/T)$ are also expected from global analysis of $B$ decays to two 
pseudoscalar mesons \cite{Chiang:2006ih}. As can be seen from Table \ref{Tab:tab2}, for all fits with free $\kappa$ and 
no constraint on $|C|/|T|$, 
$Arg(C/T) \sim -2.7 \approx 205\degree$, making our results consistent with the earlier $B\to \pi K$ fits. 
For the constrained SM-like fits, $Arg(C/T) \sim -2.35 \approx 225\degree$.

\begin{figure}
	\centering
	\subfloat[$\delacp$]{\includegraphics[width=0.75\linewidth]{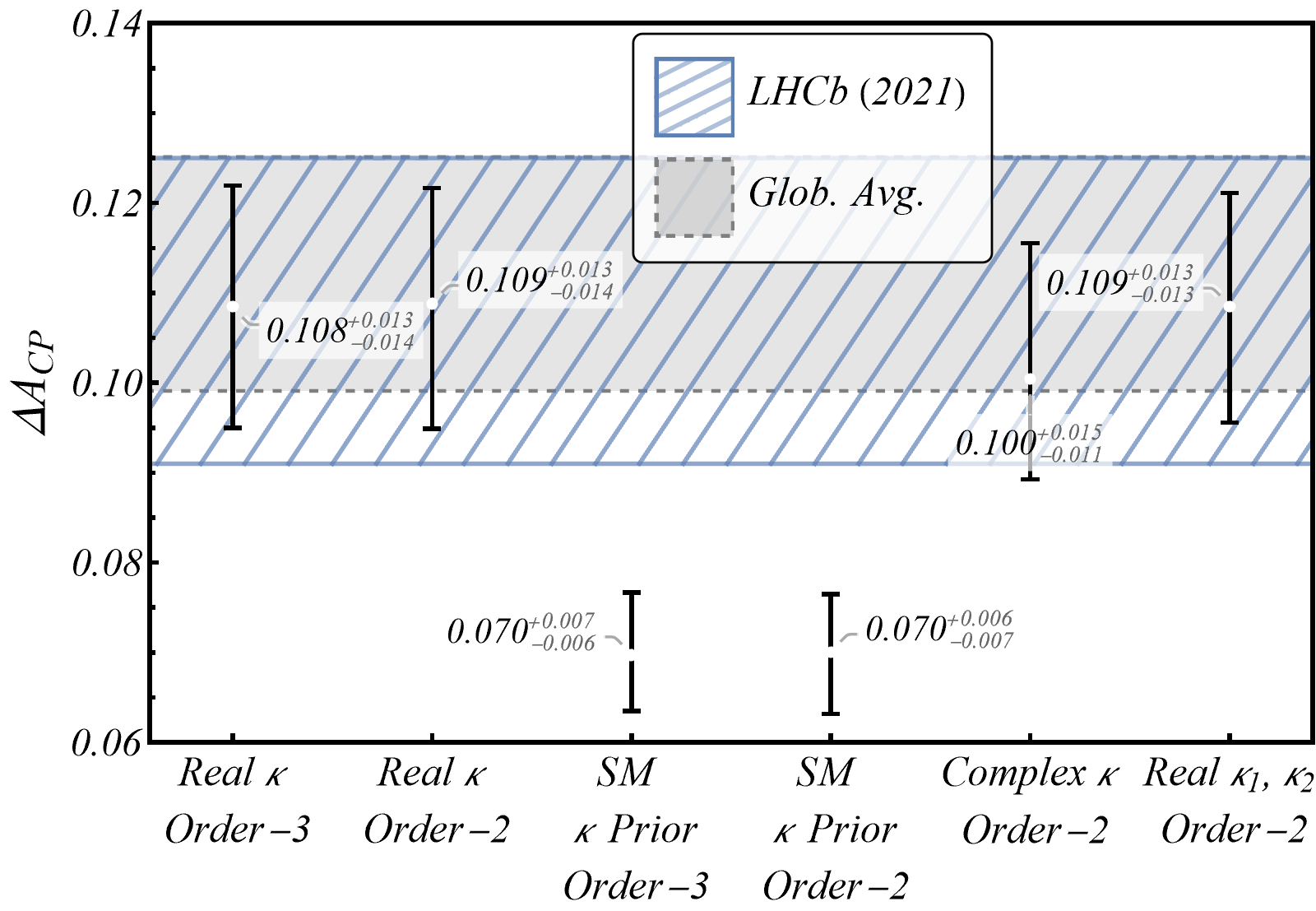}\label{fig:pred1d1}}\\
	\subfloat[$\Delta_4$]{\includegraphics[width=0.75\linewidth]{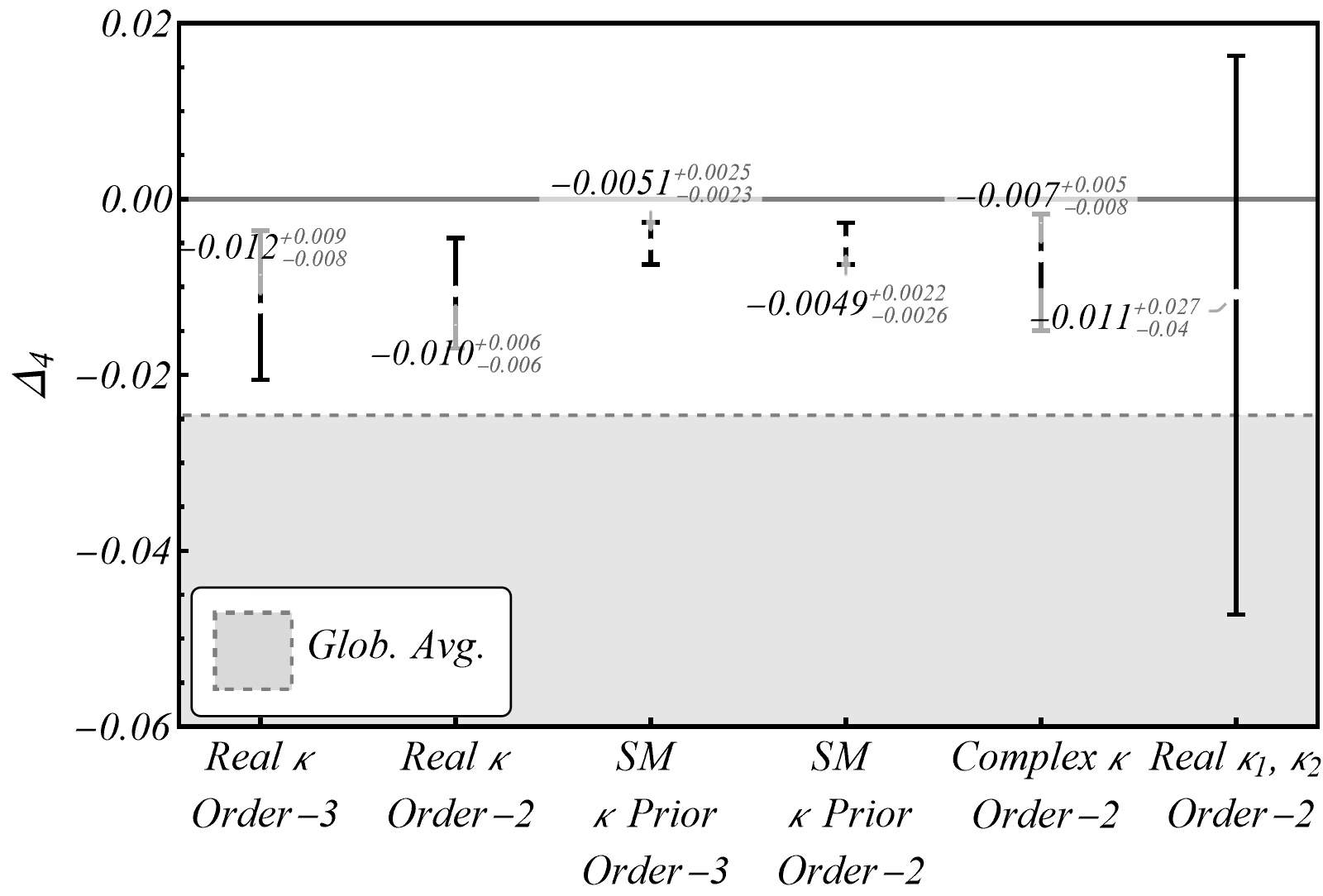}\label{fig:pred1d2}}
	\caption{\footnotesize The predicted 1D distributions of observables $\delacp$ and $\Delta_4$. 
		Figure \ref{fig:pred1d1} compares various predictions of $\delacp$ with experiment, while Figure \ref{fig:pred1d2} does the same for $\Delta_4$.}
	\label{fig:pred1d}
\end{figure}

\begin{figure}
	\centering
	\subfloat[$\delacp$ -- $\Delta_4$]{\includegraphics[width=0.49\linewidth]{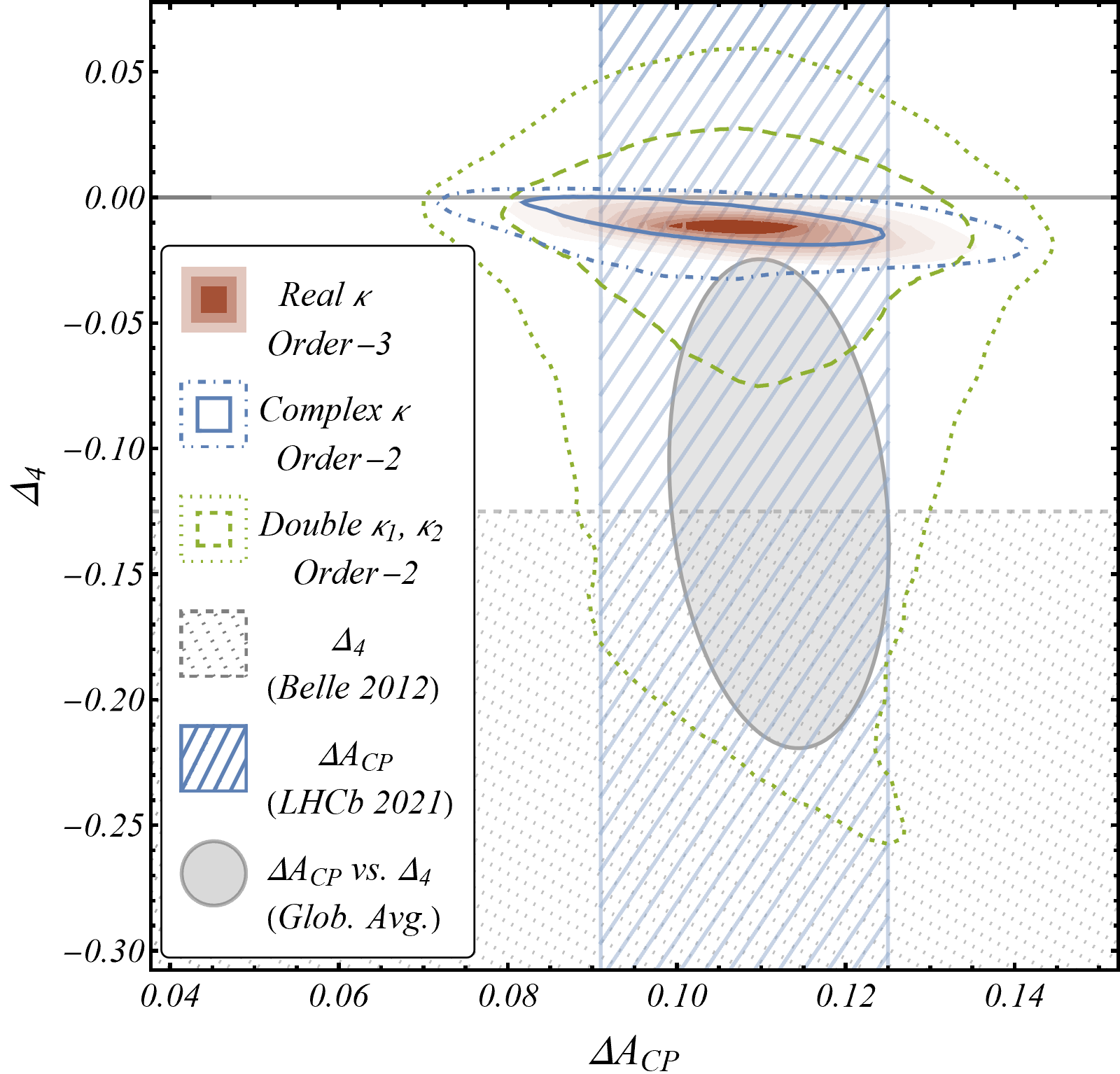}\label{fig:pred2d1}}~~
	\subfloat[$\delacp$ -- $\Delta_4$ zoomed in]{\includegraphics[width=0.49\linewidth]{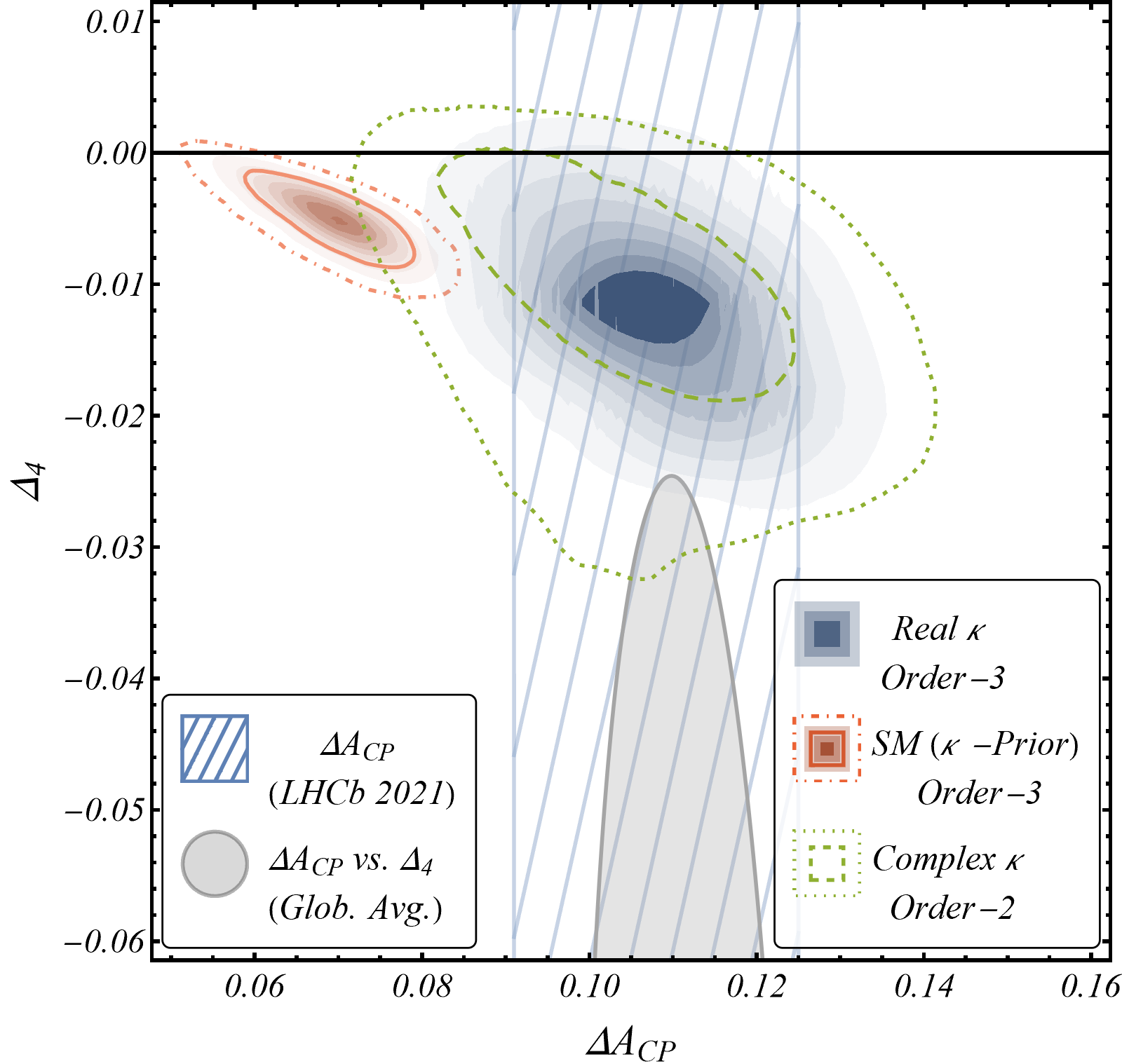}\label{fig:pred2d2}}
	\caption{\footnotesize The predicted combined distributions of observables $\delacp$ and $\Delta_4$. 
	Fig.\ \ref{fig:pred2d1} compares the experimental results with predicted combined distributions in the 
	$\delacp$--$\Delta_4$ plane for various fit results. Fig.\ \ref{fig:pred2d2} zooms in to the predicted distributions 
	from several fits. The first (second) figure exclusively contains the \textit{Double $\kappa$ Order-2} 
	(\textit{SM-like $\kappa$-Prior Order-3}) fit. All experimental bands and error ellipses denote only $1\sigma$ 
	confidence intervals.}
	\label{fig:pred2d}
\end{figure}

\subsection{Predictions}\label{sec:predictions}

Before we talk about predictions of observables, namely, $\delacp$ and $\Delta_4$, 
the reader may note that the global average of the observable $\delacp$, defined in Eq.\ (\ref{eq:delACP}) and quoted in the 
introduction~\cite{Aaij:2020wnj}, does not contain the latest results from Belle-II, as the latter results came after the publication of 
Ref.~\cite{Aaij:2020wnj}. Also, the only measurement of the observable $\Delta_4$, defined in Eq.\ \ref{eq:del4}, can be found in 
Ref.~\cite{Duh:2012ie}, which is almost a decade old. We have thus used the results listed in Table \ref{Tab:tab1} and 
found the global averages of the observables and in turn, that of $\delacp$ and $\Delta_4$. The results are,
\beq
\label{eq:globavgUs}
\delacp^{\rm global}(\pi K) = 0.112 \pm 0.013\,, \ \ \ \ 
\Delta_4^{\rm global}(\pi K) = -0.122 \pm 0.097\,,
\eeq
with correlation $= -0.175$. From here onwards, whenever we mention the global average of these observables, we will mean these 
numbers.

We have mentioned earlier that $\delacp$ is expected to be zero in SM~\cite{Gronau:1998ep} and the main goal of 
the present work is to check the robustness of that claim. The rationale for this claim is the expected smallness 
of $C$ and the expected small value of $Arg(C/T)$. Once these two assumptions are relaxed, which still does not 
take us beyond the SM, $\delacp$ need not be small.
As have been shown in the earlier sections, the favoured 
parameter spaces for most of the parameters are consistent with their respective SM expectations within $2\sigma$.
Using the extended sample of the posteriors 
of these fits, we can find the predicted distributions of $\delacp$ and $\Delta_4$. 
Fig.~\ref{fig:pred1d1} compares the predicted values of $\delacp$ with the global average, as well as 
the recent LHCb measurement~\cite{Aaij:2020wnj}. 
We have used the mode (\textit{maximum a posteriori}) and $1\sigma$ high-density CIs around them in showing the predicted distributions of the observables.
We see that both real $\kappa$ (unconstrained) fits yield very similar values of $\delacp$. 
The complex $\kappa$ fit provides slightly shifted, but completely consistent results. 
For the SM-like fits with $\kappa$ as prior, $\delacp$ is smaller than the global average. 
The central result is that the predictions for these SM-like fits are more than $5 \sigma$ away from $0$, but are 
consistent with both the LHCb result and the global average within $2 \sigma$. This shows that the data, 
though not completely consistent with the SM yet, is nowhere near the assumed large deviation of $\sim8 \sigma$.

The case of $\Delta_4$, as seen in Fig.~\ref{fig:pred1d2}, is more interesting. First of all, except the 
Double-$\kappa$ Order-2 fit, none of the predictions are consistent with either zero or the global average 
within $1\sigma$, though they are quite close to that. The biggest 
uncertainty comes from the Double-$\kappa$ fit. This prediction contains both the high-probability 
regions of the corresponding posterior. 

Fig.\ \ref{fig:pred2d} shows the correlation between $\delacp$ and $\Delta_4$. One may note that the global average ellipse 
is only at $1\sigma$, so none of the fits show any serious tension. However, if a more precise measurement unambiguously
points to a negative definite value of $\Delta_4$, the Double-$\kappa$ fit will be favoured, which definitely indicates NP.
Thus, we conclude that
along with $\delacp$, $\Delta_4$ too may act as an indicator for NP, or some new SM dynamics. 
Moreover, as can be seen from the varying uncertainties of the predictions in Fig. \ref{fig:pred1d2}, it  
may differentiate among all these possibilities.

\section{Discussion}\label{sec:summary}

In this paper, we have critically analysed the data on all the $B\to \pi K$ modes coming from all experiments, and 
checked how far the claim of an $8\sigma$ tension between the global average of $\Delta A_{\rm CP}$ and its
SM expectation can be sustained. Thanks to all the independent experiments, there is no scarcity of data, 
and the fitting procedure, as outlined previously, makes sense. What we find is more or less on the expected line:\\
(i) If we take a very naive estimate of the SM topological amplitudes, as dictated by the CKM elements, there is 
no reasonable fit with the data, be it Bayesian or frequentist.\\
(ii) Even within the framework of the SM, the colour-suppressed tree amplitude, $C$, may be significantly larger than the 
naive prediction. If we extend the allowed region for $C$, keeping $|C|/|T|\le 0.5$, we are still in the SM-like region, but the fit is considerably improved; in fact, one obtains a perfectly acceptable fit, and the fitted value of $\Delta A_{\rm CP}$  is within $2\sigma$ of the global average when the colour-suppressed tree amplitude is allowed to vary in the above mentioned range.
We have checked the allowed region with several fits,
neglecting and including suppressed contributions, and playing with the relative importance of the electroweak
penguin WCs. Everything gives the same result: the posterior distribution is almost identical, and the suppressed 
amplitudes are hardly constrained. 

Thus, the first conclusion is that there is no immediate need to go beyond the SM, although a more precise estimate of
various amplitudes  is welcome. There is now a tension of about $2\sigma$ between the global average and the 
best-fit value of $\delacp$; with more precise data this tension can grow and that will be a serious indication for beyond-SM
dynamics.

There is a second conclusion, too. For the double-$\kappa$ fit, the
parameter space shows another best-fit region, which, in fact, contains
the global maximum-likelihood estimate.
However, the region is far from what is allowed by the SM. For this region, NP is definitely
indicated. Thus, we have found that both the observables, namely, $\delacp$ and $\Delta_4$, 
can differentiate between these two high-probability regions 
of parameter space, or, in other words, act as smoking guns for new physics. We, therefore,
urge our experimental colleagues to measure this quantity as precisely as possible. 

 \acknowledgements
The authors acknowledge the Science and Engineering Research Board, Government of India, for the 
grants CRG/2019/000362 (AK and SR), MTR/2019/000066 and DIA/2018/000003(AK).

\appendix*
\section{Bayesian Terminology}\label{sec:app_Terms}
For a point or interval estimation of a parameter $\theta$ in a model $M$ based on data $y$, 
Bayesian inference is based off the Bayes' Theorem:
\begin{align}
	p(\vec{\theta}|y) = \frac{p(y|\vec{\theta})~p(\vec{\theta})}{p(y)} \propto p(y|\vec{\theta})~p(\vec{\theta})\,,
\end{align}
where 
\begin{enumerate}
	\item $p(\vec{\theta})$ is the prior probability density (in short, \textit{prior}) for the parameter-vector 
	$\vec{\theta}$, encapsulating all of our initial knowledge about the parameters. In this work, the priors for the 
	free parameters are set as uniform distributions of a very large range, whereas the theoretical inputs are 
	incorporated as multi-dimensional Gaussian distributions.
	\item $p(y|\vec{\theta})$ is the `likelihood function' (in short, \textit{likelihood}). It quantifies the likelihood that the 
	observed data would have been observed as a function of $\vec{\theta}$ (but it is not a probability density for 
	$\vec{\theta}$).
	\item $p(y)$ is the evidence, defined as $p(y) = \int p(y|\vec{\theta})~p(\vec{\theta})~d\vec{\theta}$ and is 
	just a constant for our purpose, {\em i.e.}, parameter estimation. 
	\item $p(\vec{\theta}|y)$ is the coveted posterior (or `inverse', in old usage) probability distribution (in short, 
	\textit{posterior}) of $\vec{\theta}$, given the data $y$. We generate samples from this distribution (actually 
	the un-normalised one, ignoring the generally intractable `evidence') by running an MCMC process. 
	\item For estimating any single parameter $\theta_j$ among $n$ such parameters, we need to find the 
	one-dimensional (1-D) marginal distribution of that parameter by integrating the full posterior over all other 
	parameters: $p(\theta_j|y) = \int p(y|\vec{\theta})~d\theta_1\ldots d\theta_{j-1}~d\theta_{j+1}\ldots d\theta_n$. 
	In practice, once the MCMC sample is generated, marginalising is as trivial as neglecting all other parameter
	values from the sample. Similarly, higher-dimensional marginal posteriors can also be generated. Such 2-D 
	marginal posteriors are used to depict the parameter spaces in most places in this work.
\end{enumerate}

\textit{Credible Intervals (CIs)}: In Bayesian parlance, the interval within which the appearance of an unobserved 
parameter value has a particular probability, is called a credible interval (credible region, for multivariate cases).

As we obtain a probability distribution (posterior) after a Bayesian analysis, point (central tendency) or interval 
(dispersion) estimation is not unique and quite problematic. The best Bayesian analogue of the MLE is the Maximum 
\textit{a posteriori} Probability (MAP) estimate, which is (a) really ambiguous for multi-modal distributions (as in 
the case of the \textit{`Real $\kappa$ Global Order-2'} fit), (b) generally uncharacteristic of the majority of the 
posterior, and (c) is not invariant under re-parametrisation. 
We have thus only mentioned medians as point-estimates of parameters, only for the unimodal \textit{`Local'} fits. 
Furthermore, the credible intervals (CIs) around these estimates, in addition to implying completely different 
conceptual things from confidence-intervals (their frequentist analogues), depict different regions with different 
probability content, in general.

This is why we refrain from using the point estimates for calculating our numerical predictions, and instead use 
the whole posterior samples to do that.

\bibliography{B2KPi}

\end{document}